\documentclass[10pt]{article}
\usepackage[english]{babel}
\usepackage{sansmathfonts}
\usepackage[T1]{fontenc}
\usepackage{amsmath,amsfonts,amssymb,amstext,amsmath,amsthm,amscd}
\usepackage{mathtools}
\usepackage[a4paper,top=2cm,bottom=2.5cm,left=2.5cm,right=2.5cm]{geometry}
\usepackage{enumitem}
\usepackage{mathrsfs}
\usepackage{textcomp}
\usepackage{braket}
\usepackage{colortbl}
\usepackage{bbm}
\usepackage{caption}
\usepackage{subcaption}
\usepackage{graphicx}
\usepackage{adjustbox}  
\usepackage{multirow}  
\usepackage{array}
\usepackage{comment}

\usepackage{breakcites}
\usepackage{mathtools}
\usepackage{authblk}

\usepackage{mathrsfs}
\usepackage{dsfont}
\usepackage{graphicx}
\usepackage{subcaption}
\usepackage{wrapfig}
\usepackage{indentfirst}
\usepackage{latexsym}
\usepackage{sidecap}
\usepackage{booktabs}
\usepackage{hyperref}

\usepackage{lipsum}
\usepackage{cleveref}
\usepackage[dvipsnames]{xcolor}
\usepackage[nodisplayskipstretch]{setspace}

\newcommand\myshade{85}
\colorlet{mylinkcolor}{violet}
\colorlet{mycitecolor}{YellowOrange}
\colorlet{myurlcolor}{RoyalBlue}

\hypersetup{
  colorlinks=true,
  linkcolor=myurlcolor,
  citecolor=mycitecolor!\myshade!black,
  filecolor=magenta,      
  urlcolor=magenta,
}

\newcommand{\marco}[1]{\textcolor{red}{#1}}

\title{Mechanistic driven TCP and NTCP modeling for particle therapy accounting for a broad range of physical irradiation  parameters and tissue environmental conditions}

\author[1,2]{Marco Battestini}
\author[1,2]{Jules Morand}
\author[1,2]{Giulio Bordieri}
\author[1,3]{Marta Missiaggia}
\author[1]{Emanuele Scifoni}
\author[1,4]{Francesco G. Cordoni}

\affil[1]{Trento Institute for Fundamental Physics and Application (TIFPA), via Sommarive 15, Trento, 38123, Italy}
\affil[2]{Department of Physics, University of Trento, via Sommarive 14, Trento, 38123, Italy}
\affil[3]{Department of Physics \& Astronomy, Louisiana State University (LSU), 202 Nicholson Hall, Baton Rouge, LA 70803, US}
\affil[4]{Department of Civil, Environmental and Mechanical Engineering, University of Trento, via Mesiano 77, Trento, 38123, Italy}

\begin{document}

\maketitle

\begin{abstract}
In conventional radiotherapy (RT), the probability of controlling tumor growth is quantified using Tumor Control Probability (TCP) models. Instead, the probability of experiencing a side effect after the irradiation of healthy tissues and organs is typically assessed using the concept of Normal Tissue Complication Probability (NTCP), an additional crucial metric for evaluating and comparing treatment plans. 

This work is dedicated to the development, implementation, and application of a general mechanistic model to describe the effects of particle therapy (PT) on different tissue organizations beyond Poissonian assumptions, extending the Generalized Stochastic Microdosimetric Model (GSM$^2$), i.e., a  stochastic radiobiological model that describes the time evolution of DNA lesions in a cell nucleus according to microdosimetric principles, to the study of macroscopic biological systems. Specifically, we extend the biological stage of radiation damage of the GSM$^2$ model to larger spatial and temporal scales, involving cell populations with a specific geometric and functional architecture.

In particular, the GSM$^2$-driven NTCP and TCP models are mechanistic-driven dose-response models that integrate the GSM$^2$ into the RSM model, allowing a comprehensive investigation of clinically relevant endpoints, such as healthy tissue complications and tumor control predictions for conventional irradiation. The model’s single-cell resolution allows it to account for energy deposition and tissue heterogeneity, considering different organ volume effects, cell type distributions, and oxygen gradients for different radiation qualities of the beam, that is, type, energy, and LET of radiation, and various fractionation schemes. We show the interplay between physical and environmental parameters on the induction of side effects on healthy tissues, for different radiation qualities and fractionation schemes, and we highlight the impact of biochemical heterogeneities in the target environment, for tumor response.
\end{abstract}


\tableofcontents
%\linenumbers

\section{Introduction}
\label{0_Introduction}
\counterwithin{figure}{section}
\setcounter{figure}{0}

The primary objective of radiotherapy (RT) is to counteract tumor progression and maintain control over tumor volume growth. To achieve this effectively, a precise understanding of how radiation dose influences tumor response is essential, and various mathematical approaches have been developed in the radiotherapy community over the last few decades to characterize this dose-effect relationship \cite{webb1993model, kallman1992tumour, niemierko1993implementation}. In particular, the response of a single clonogenic cell, i.e., capable of proliferating, to ionizing radiation (IR) can be described through the Linear-Quadratic (LQ) model, where the DNA molecule is the critical target, whose outcome is the cell survival probability. In this way, the Tumor Control Probability (TCP) for a given tumor volume with $N^{cell}$ cells, which are characterized by a specific radio sensitivity $\alpha/\beta$, can be defined as a function of the delivered dose $D$, representing the probability that all clonogenic cells die \cite{webb1993model}.

The second aim of radiation treatment is to spare as much of the surrounding healthy tissue as possible \cite{de2019radiotherapy, citrin2026effects}. However, the study of the radiation response of macroscopic biological structures, such as tissues or organs, is complicated because different types of cells may be present, characterized by different radiosensitivities. Furthermore, different organs exhibit distinct responses to radiation. Some organs lose function when even a small portion of their tissue is damaged, whereas others are impaired only when nearly all constituent cells are inactivated. This phenomenon, commonly referred to in the literature as the volume effect, is typically modeled through smaller aggregations of cells known as functional sub‑units (FSUs) \cite{withers1988fsu, bidanta2025functional, citrin2026effects}. FSUs represent biological structures with defined organization and function (e.g., lung alveoli or nephrons) and constitute one of the most widely adopted modeling frameworks for describing organ‑level radiation response. In addition, the FSUs can be organized in different structural arrangements, forming serial and parallel tissues. In a serial arrangement, the functional sub-units are connected as the links of a chain, while in a parallel organization, the FSUs are independent of each other. For this reason, in the case of a fully serial organ, if a single FSU is irreparably damaged by IR, this event can trigger the loss of organ functionality, leading to a complication, such as for the spinal cord or the esophagus. Instead, for a parallel organ, the inactivation of a single FSU does not disturb the correct functionality of the organ, such as for the liver, the lung, or the parotid glands; in this case, the inactivation of FSUs above a specific threshold \cite{scholz2006dose}, quantifiable by a minimum number of preserved FSUs, can cause a complication. 

Therefore, tissue architecture, i.e., the different FSUs arrangements, can explain the different dose-volume dependencies of radiation response, namely how an organ responds to partial irradiation of its volume \cite{withers1988treatment}. In this context, we can define the volume effect as the way in which an organ or a tissue responds to damage after irradiation, depending on its structural organization \cite{battestini2022including, niemierko1999geud, kallman1992tumour}. In particular, serial organs are characterized by what is typically referred to as a \textit{small volume effect}, because the inactivation of a single FSU, and thus of a small irradiated volume, can lead to organ dysfunction. Instead, for parallel organs, the volume effect is large, that is, the organ can tolerate a certain number of damaged FSUs. In practice, real complex organs are organized as a combination of serial and parallel architectures, like for the nephron structure. Furthermore, the same organ can develop several different complications, responding in some cases as a serial organization, while in others as a parallel architecture. The typical example is rectal bleeding compared to diarrhea.

The notable improvement of radiotherapy techniques and treatment planning systems (TPSs) in recent years \cite{citrin2026effects} has led to the technical possibility of generating a wide variety of treatment plans and dose distributions, which must be compared in a simple and fast way. Several studies on radiation-induced normal tissue toxicity have been performed, particularly for standard RT with photons \cite{emami1991tolerance, burman1991fitting}. The probability of encountering a side effect has been quantified by proposing the concept of Normal Tissue Complication Probability (NTCP), a key quantity to classify the goodness of treatment plans. In doing this, however, it is necessary to take into account possible physical and tissue environmental heterogeneities, such as in energy deposition and dose distribution, chemical and biological properties of the irradiated volume, etc. In addition, there are also possible practical difficulties in mathematical modeling, such as patient variability, different radiation complications for the same type of organ, different clinical methods for the evaluation of the endpoint, and additional clinical factors (surgery, chemotherapy, co-morbidities, etc.). All of this led to an intensive modeling study of the NTCP, using different approaches, namely: (i) Statistical models, which are based on the correlation between dosimetric or clinical variables and a specific toxicity endpoint, using logistic regression analysis. However, they are characterized by few or no biological considerations, such as the univariate (UVA) and multivariate (MVA) logistic analyses \cite{fellin2009clinical}; (ii) Mechanistic models, which are based on tissue organization and cellular radiosensitivity, such as the Relative Seriality Model (RSM) \cite{kallman1992tumour}. Although they are the most mathematically grounded models, their clinical use is limited; (iii) Phenomenological models, which are based on a sigmoidal dose-response curve and are the most widely used approach, such as the Lyman–Kutcher–Burman (LKB) model \cite{kutcherburman1989ntcplkb}.

In particular, the RSM is a common NTCP model, which was originally proposed by Kallman et al. in \cite{kallman1992tumour}. This is a more radiobiologically based NTCP model, compared to phenomenological models such as the LKB model \cite{kutcherburman1989ntcplkb}. In fact, it takes into account the architecture of the organ through the definition of a relative seriality parameter $s$. Instead, in the context of the RSM \cite{kallman1992tumour}, the tumor is parameterized as a purely parallel organ. Thus, the TCP is defined as the probability of killing the tumor. 

However, the RSM, like most of the NTCP models in literature, has been developed specifically for photon radiotherapy and lacks a robust radiation biophysical basis, neglecting tissue heterogeneity and key radiation-induced stochasticities, affecting energy deposition, damage formation, and their temporal evolution. This missing analysis hinders the effective usage of this approach to the case of particle therapy (PT), where all of these effects are known to play a crucial role. Thus, the main goal of this work is to develop a new mechanistically grounded model that, starting from a rigorous mathematical formulation of particle-induced radiobiological effects and incorporating principles derived from the RMS framework, introduces a novel TCP/NTCP mechanistic approach. This model enables the investigation of particle therapy outcomes—from single‑cell responses to whole‑organ endpoints, under a wide range of tissue characteristics and irradiation conditions.

A crucial complication in generalizing NTCP from conventional radiotherapy to particle therapy (PT) is how to incorporate radiation quality into such a model. Radiation quality characterizes the radiation field and is typically described by macroscopic quantities such as particle type, energy [MeV/u], and linear energy transfer (LET) [keV/$\mu$m], defined as the energy transferred per unit path length. While LET provides a practical macroscopic descriptor, it does not capture the stochastic nature of energy deposition at the microscopic scale. A more complete detailed description is offered by microdosimetry \cite{Ros}, which accounts for the full energy spectrum and stochastic effects in energy deposition through random variables rather than mean values \cite{Bellinzona2021}. The fundamental microdosimetric quantities are the \textit{specific energy} $z=\frac{\varepsilon}{m}$ [Gy], where $\varepsilon$ is the energy imparted in matter of mass $m$, and the \textit{lineal energy} $y=\frac{\varepsilon_1}{\bar{l}}$ [keV/$\mu$m], where $\varepsilon_1$ is the energy imparted by a single deposition event and $\bar{l}$ is the mean chord length \cite{Bellinzona2021}. Bridging the macroscopic and microscopic perspectives, the \textit{amorphous track} (AT) structure approximation \cite{kase2007biophysical} provides an effective framework for modeling the spatial pattern of energy deposition, replacing the stochastic structure of particle tracks with a radially symmetric local dose distribution around the trajectory. The AT formalism underlies the two main radiobiological models in PT: the Microdosimetric Kinetic Model (MKM) \cite{kase2007biophysical} and the Local Effect Model (LEM) \cite{pfuhl2022comprehensive}, and is adopted here as the basis for linking radiation quality to biological response. Such a biological response is typically quantified in PT through the Relative Biological Effectiveness (RBE), defined as the ratio of the reference radiation dose, typically photons, to the ion dose required to achieve the same biological effect \cite{scholz2003effects}. RBE depends on physical parameters such as absorbed dose and radiation quality, as well as biological factors such as cell line. Although radiation quality, through its role in shaping the microscopic spatial pattern of energy release, strongly influences macroscopic endpoints such as NTCP and TCP, it is rarely incorporated at this scale. This is primarily due to the computational cost involved: accurate modeling requires micrometer-scale resolution or cell-by-cell calculations, which have severely limited the development of mechanistic models capable of describing macroscopic endpoints across different particle types and energy ranges.

In this work, we propose a novel \textit{mechanistic, probabilistic} TCP/NTCP framework that integrates the Generalized Stochastic Microdosimetric Model (GSM$^2$)~\cite{cordoni2021generalized, cordoni2022cell, cordoni2022multiple} into the RSM formalism. This integration enables a unified and consistent description of tumor response and healthy-tissue complications for conventional irradiations, including volume effects, across a wide range of physical, biological, and environmental conditions. GSM$^2$ is a fully probabilistic biophysical model that describes the formation and evolution of DNA damage from microdosimetric principles~\cite{Bellinzona2021, missiaggia2020microdosimetric, missiaggia2021novel, missiaggia2023investigation}, and has been validated and extended in several recent studies~\cite{Bordieri2024, Battestini2023, battestini2024multiscale, Missiaggia2024, bordieri2025integrating}.

Here, we develop the mathematical formulations of the mechanistically driven NTCP and TCP models and describe their numerical implementation. We then present the main predictive results for complication probabilities and tumor-control likelihood across different irradiation settings. Our framework allows us to investigate how oxygen distribution, tissue seriality, fractionation schemes, and radiation quality influence the macroscopic biological outcome.

A key contribution of this work is the introduction of a general probabilistic dose--effect framework capable of modeling NTCP and TCP with different volume effects for a broad spectrum of particle types and energies. This is achieved through an efficient single-cell--resolution computation of radiation-induced biological damage, scalable from the cellular level up to millimeter-scale tissue structures. Beyond radiation quality and cellular dose---modeled via the AT microdosimetric formalism, our approach explicitly incorporates tissue heterogeneity, allowing each cell to be characterized by distinct biochemical environments (e.g., oxygenation levels) and radiosensitivity parameters.

Overall, we demonstrate how both the physics of the radiation field and the heterogeneous structure of living tissue, combined with volume effects, can give rise to a rich variety of biological responses, spanning from the single-cell to the organ level.

%% MATERIALS AND METHODS
\section{Materials and methods}
\label{1_Materials_Methods}

\begin{figure}[htbp]
    \centering
    \includegraphics[width=0.7\linewidth]{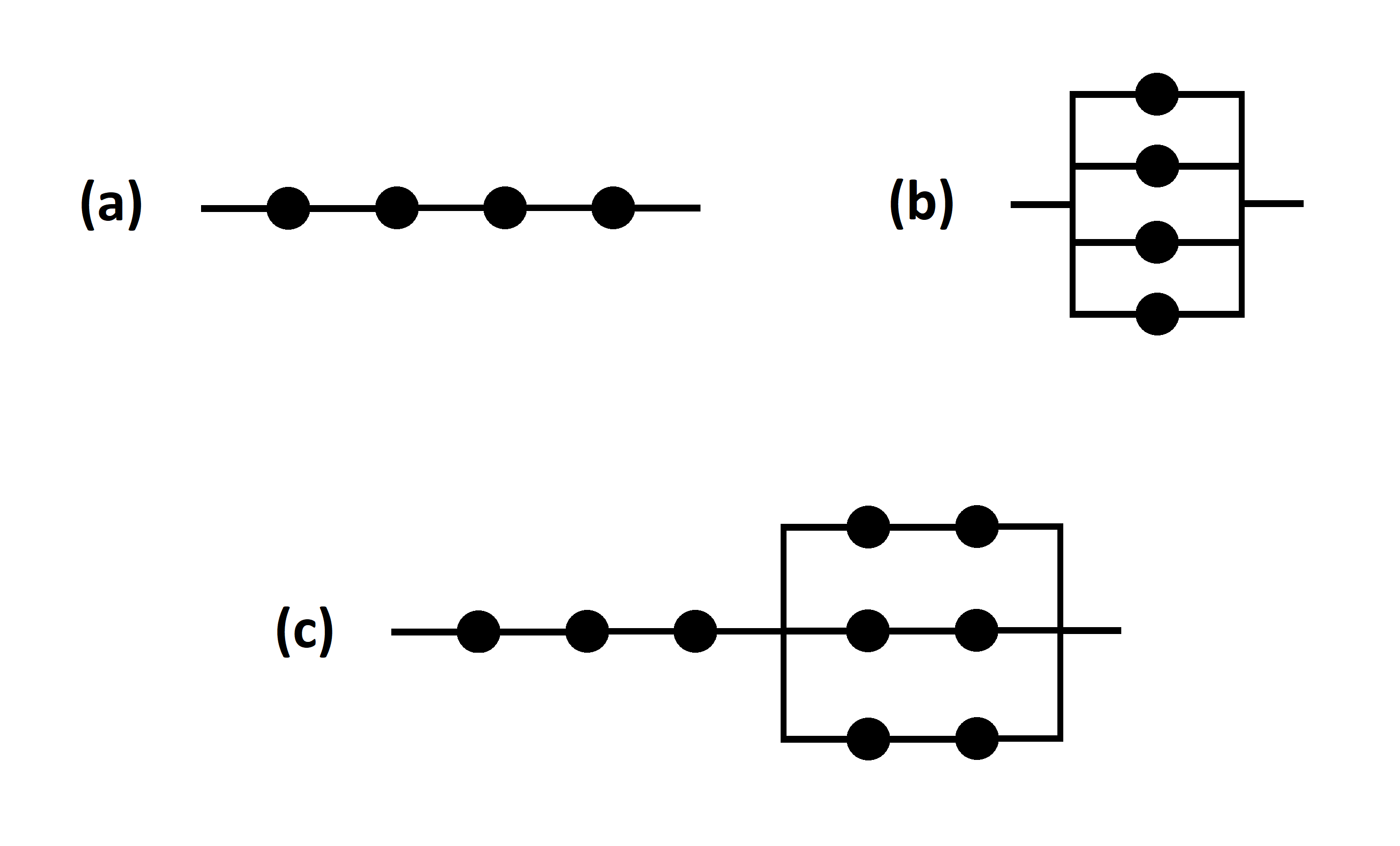}
    \caption{Schematic representation of \textbf{(a)} serial, \textbf{(b)} parallel, and \textbf{(c)} complex organ structure. The points represent the functional sub-units (FSUs), while the solid lines represent the connections between FSUs.}
    \label{fig:organs_FSU}
\end{figure}

\begin{figure}[htbp]
    \centering
    \includegraphics[scale=0.6]{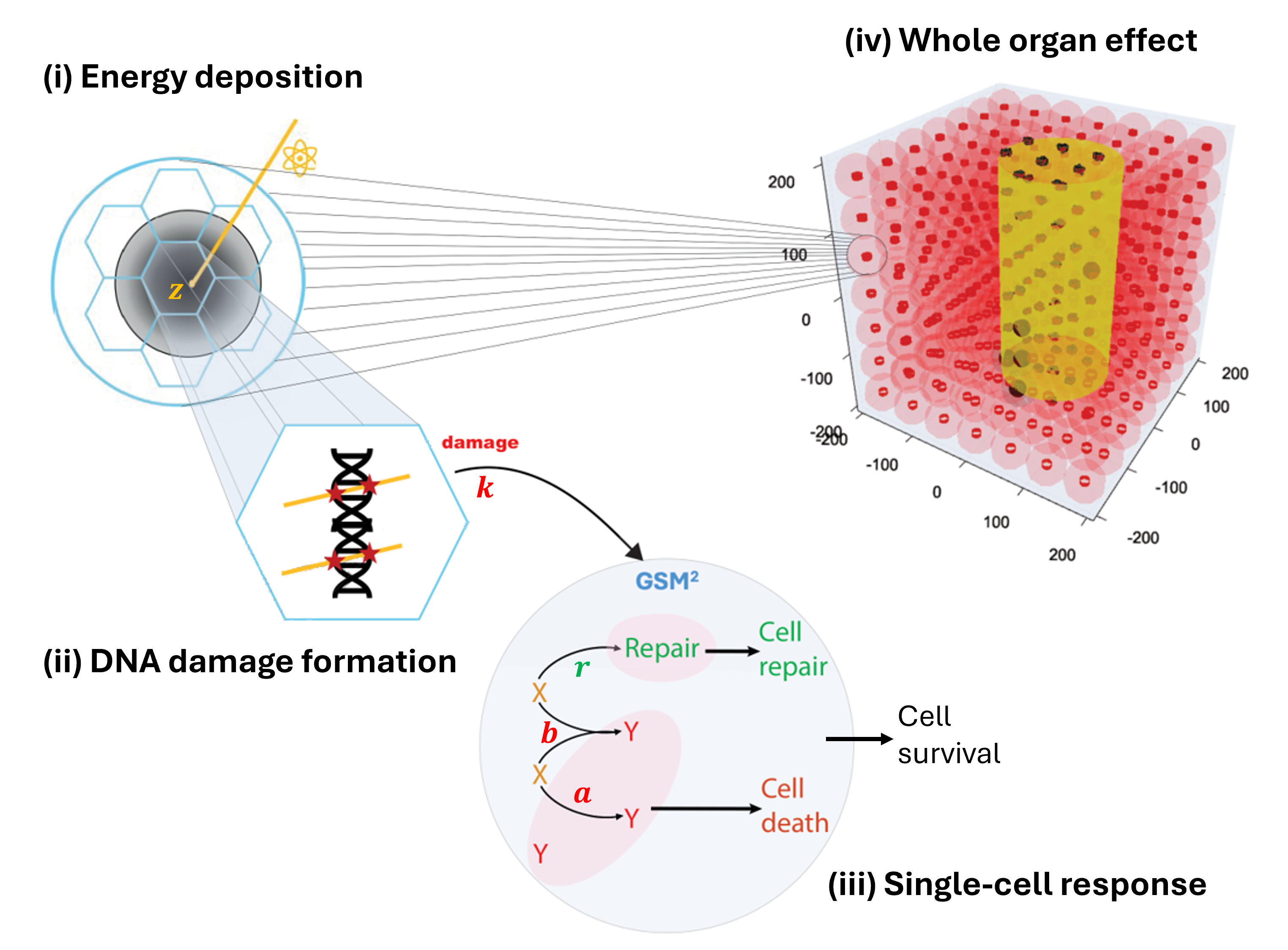}
    \caption{Schematic representation of the GSM$^2$-driven NTCP and TCP workflow. The model takes into account different spatial and temporal scales of radiation-induced biological damage, which are (i) the physical stage (energy deposition), (ii) the biochemical stage (DNA damage formation), and (iii-iv) the biological stage (single-cell response and whole organ/tumor effect). The physical stage describes the energy deposition on a cell nucleus according to microdosimetry. The bio-chemical stage encompasses radiation-induced sub-lethal and lethal DNA damages, incorporating modulation effects driven by radiation quality and oxygen availability. The biological stage models the temporal evolution of DNA lesions through three distinct biological pathways, i.e., repair, death, and combined-death cellular mechanisms, and the macroscopic response of the organ and tumor volume, considering its specific biological architecture, which is described by the seriality parameter $s$.}
    \label{fig:GSM2_scheme_NTCP}
\end{figure}

We consider a cell population of $N^{cell}$ cells, which are considered spheres of radius $R_{cell}$, as illustrated in Figure \ref{fig:GSM2_scheme_NTCP}. The cells are arranged on a 3D grid of the target volume, with the centers of the cells spaced $2R_{cell}$ apart from each other. In agreement with standard modeling literature in particle therapy \cite{kase2007biophysical}, the cell has a cylindrical nucleus, which is the target of the GSM$^2$ model, of radius $R_N$ in the $xy$-plane and length $2\,R_N$ in the $z$-direction, namely the projectile direction. The nucleus is divided into $N_d$ mathematical domains, following that proposed in \cite{cordoni2021generalized}. In particular, the cylinder is divided into $N_z$ circular layers, and each circle is subdivided into $N_{xy}$ circular domains, with a typical radius $R_d$ in the order of micrometers. Thus, the cylinder is composed of a total of $N_d = N_z \cdot N_{xy}$ independent domains $d$. The subdivision into domains allows for consideration of physical, chemical, and biological spatial information. In fact, in this way, it is possible to have a discretized spatial distribution of energy at the micrometer scale, and take into account these energy distributions in the local onset of DNA damage. 

The GSM$^2$-driven TCP/NTCP developed model is a modular sequence of four stages of radiation-induced damage. The processes that lead to radiation-induced organ response through the formation of DNA lesions start with (i) the physical stage, which is characterized by energy deposition by the primary ionizing projectile hitting the surface of the cell nucleus. The deposition event is extremely fast and is considered instantaneous for this model. Then, the energy deposited is spatially distributed in the horizontal plane of each cell nucleus. This dose deposition process is followed by (ii) the biochemical stage. During this stage, DNA damage is caused by the action of IR, which can be modulated by the radiation quality, i.e., particle type, energy, and LET, and the presence of molecular oxygen. The last step is the biological stage, which describes (iii) the single-cell response and (iv) the whole organ effect. In particular, the DNA lesions can undergo three different possible biological pathways, such as the repair process or two different death events, leading to the probability that cells survive after irradiation, which is the final biological endpoint of the GSM$^2$ model. Lastly, the developed model captures the macroscopic response of tumors and organs, depending on their specific biological architecture and volume effect. The first three steps (i)–(iii) were already developed in previous research \cite{cordoni2021generalized, cordoni2022cell, Battestini2023, battestini2024multiscale}, whereas module (iv), together with its efficient integration with the existing components, constitutes the main novelty of the present work. It is worth noting that the inclusion of step (iv) requires significant computational enhancements to the earlier modules to ensure that the overall workflow remains efficient.

In the following paragraphs, we particularly focus on the description of the mathematical framework of the model for describing the energy deposition, DNA damage formation and evolution, and the organ's response.

\subsection{The energy deposition model}
\label{Subsubsection: physics}

Depending on the considered application, a region of the cell population, which can also be the entire cell population as illustrated in Figure \ref{fig:GSM2_scheme_NTCP} (iv), is irradiated with a circular beam of radius $R_{beam}$ with a certain LET spectrum, which characterizes the radiation quality, and dose $D$. The beam penetrates the tissue along the $z$-direction. 

The number of incident particles is randomly sampled from a Poisson distribution of average fluence obtained by inverting the dose–LET relation, that is
\begin{equation}\label{EQ:Nparticle}
N_p = \frac{\pi R_{beam}^2 \cdot \rho \cdot D}{1.6 \cdot 10^{-9} \cdot \mathrm{LET}} \,,
\end{equation}
where $\rho$ is the water density.

Each particle is randomly distributed along the planar $xy$-plane inside the beam. Also, the corresponding LET is sampled from the LET spectrum for each particle. Thus, sampling the impact position of the primary particles and the LET gives an energy stochastic deposition described by microdosimetry principles \cite{Bellinzona2021}. Therefore, each cell will experience a different energy deposition, as illustrated in Figure \ref{fig:GSM2_scheme_NTCP} (i). 

The absorbed energy of every energy deposition event $z$ is distributed on each cell nucleus according to an AT model. Following analogy with the MKM \cite{kase2007biophysical}, which represents the original inspiration of the GSM$^2$, we specifically adopted the Kiefer–Chatterjee (KC) model \cite{kiefer1986model, chatterjee1976microdosimetric}, which provides an analytical approximation of the radial dose distribution for a given projectile. The AT model postulates energy deposition by a charged particle with cylindrical symmetry, with a uniform energy distribution across a circle of radius $R_c$ centered at the impact point. This distribution diminishes proportionately to $r^{-2}$, where $r$ represents the distance from the impact point. The model assumes no energy is deposited beyond a specific distance, denoted as $R_p$. The radial energy distribution is thus computed according to the following analytical formula:
\begin{equation} \label{EQ:ATdose}
    D(r) =
    \begin{cases}
    \frac{1}{\pi R_c^2} \left( \frac{\mathrm{LET}}{r} - 2 \pi K_p \ln \left( \frac{R_p}{R_c} \right) \right) & r \leq R_c \\
    \frac{K_p}{r^2} & R_c < r \leq R_p \\
    0 & r > R_p
    \end{cases} \,,
\end{equation}
where $r$ is the radial distance from the center of the track, $K_p = 1.25 \cdot 10^{-4} \left( z^* / \beta_{ion} \right)$, $z^*$ is the effective charge, $\beta_{ion}$ is the velocity relative to the light velocity, $R_c = 0.0116 \cdot \beta_{ion}$ is the radius of the track core centered on the impact point, $R_p = 0.0616 \cdot (E/A)^{1.7}$ is the maximum radius of the track penumbra, which depends on the energy per nucleon $E/A$ of the primary radiation, $\mathrm{LET}$ is the linear energy transfer of the primary particle, which characterizes its radiation quality, and $\rho$ is the water density. Thus, in this way, we can know the absorbed dose at each domain in the $xy$-plane inside each cell nucleus, as initially proposed in \cite{Battestini2023}. %Thus, the AT model considers a uniform dose distribution in the core of the particle track, while the radial dose decreases proportionally to $r^{-2}$ in the penumbra region. 

For low-LET radiation, we assume the so-called \textit{track-segment condition} within each nucleus, meaning that the energy of the incoming particle is high enough that energy loss inside the cell nucleus can be disregarded. Thus, the LET is constant within the cylinder of length $2R_N$, in analogy to \cite{kase2007biophysical}. On the contrary, energy loss is accounted for by different layers of cells along the penetration direction $z$ of the beam. This can be computed analytically for ions, according to the Bethe-Bloch formula \cite{ICRU2016}. Knowing the energy loss by ions per unit of path length in the $z$-direction, it is possible to compute the corresponding radial dose distribution in the $xy$-plane for each layer. The total energy deposited in each cell is thus calculated, summing up all the energy deposition by the single particle.

\subsection{The DNA damage formation model}
\label{Subsubsection: DNA damage}

The model, as reported in Figure \ref{fig:GSM2_scheme_NTCP} (ii), assumes the creation of two possible types of radiation-induced DNA damage, $X$ and $Y$, which are the sub-lethal and the lethal lesions, respectively, according to the GSM$^2$ model \cite{cordoni2021generalized}. In detail, each single energy deposition $z$ can induce a random number $Z_{X}^{(j,d)}$ and $Z_{Y}^{(j,d)}$, of sub-lethal or lethal lesions, respectively, on each domain $d=1,\dots, N_d$ of the $j$-th cell. In particular, the random variables $Z_{X}^{(j,d)}$ and $Z_{Y}^{(j,d)}$ are defined as Poisson--distributed with mean
\begin{equation}\label{eq:kDNA}
    \kappa\!\left( \mathrm{LET}, [\mathrm{O_2}] \right)\, z
    = \frac{\mathrm{DSB}(\mathrm{LET})}{\mathrm{OER}\!\left( \mathrm{LET}, [\mathrm{O_2}] \right)} \, z \, ,
\end{equation}
where $\mathrm{DSB}(\mathrm{LET})$ denotes the analytical expression for the average yield of DNA double-strand breaks, calibrated from track--structure simulations~\cite{kundrat2020analytical}. Previous studies have shown that coupling the stochastic nature of particle traversals with the microdosimetric stochasticity of local energy deposition $z$ \cite{cordoni2022multiple, cordoni2022cell, missiaggia2022cell, Battestini2023, battestini2024multiscale}, together with a Poissonian description of damage induction conditioned on a given energy release, results in a \emph{richer and more realistic} probability distribution than that obtainable from purely Poisson statistics alone.

As reported in equation \eqref{eq:kDNA}, we further consider the so-called oxygen fixation effect in the induction of DNA damage, namely the increasing radioresistance of cells to a decrease in oxygen concentration. This effect is parameterized through the oxygen enhancement ratio (OER), i.e. the ratio of doses at a given oxygenation level $D (p\mathrm{O_2})$, i.e., $\le 21~\%$ $\mathrm{O_2}$, as compared to the normoxia condition $D (21~\%)$, which is oxygen partial pressure in air, i.e., $21~\%$ $\mathrm{O_2}$, to obtain the same given biological effect. Specifically, the oxygen fixation effect can be described according to 
\begin{equation}\label{EQ:OER_LET_O2}
\mathrm{OER} \left( \mathrm{LET}, [\mathrm{O_2}] \right) = \frac{K_{\mathrm{O_2}} \cdot  \frac{K_{\mathrm{LET}} \cdot M + \mathrm{LET}^\gamma}{K_{\mathrm{LET}} + \mathrm{LET}^\gamma} + [\mathrm{O_2}]}{K_{\mathrm{O_2}} + [\mathrm{O_2}]} \,,
\end{equation}
where $M$ quantifies the maximum effect, $K_{\mathrm{O_2}}$ denotes the flex point for $[\mathrm{O_2}]$, as described in \cite{epp1972radiosensitivity}, $K_{\mathrm{LET}}$ represents the flex point for $\mathrm{LET}$, that is, the average energy deposited per unit path length, and $\gamma$ is an additional fitted parameter, as reported in \cite{Scifoni2013}. The $\mathrm{OER}$ illustrates a well-established experimental phenomenon in radiation biophysics: as oxygen concentration increases, so does damage fixation, enhancing radiation lethal effects \cite{Scifoni2013}. The $\mathrm{OER}$ depends strongly on radiation quality, i.e., on $\mathrm{LET}$. In particular $\mathrm{OER}$ decreases with increasing $\mathrm{LET}$ \cite{Scifoni2013}. This should be due to the increase in the densities of the track with increasing $\mathrm{LET}$.

\subsection{The single-cell response model}
\label{Subsection: surv prob}

Given the sub-lethal and lethal lesions $X^{(j,d)}$ and $Y^{(j,d)}$, respectively, distributions within each domain $d$ of each cell $j$, the repair and death pathways can be determined as described by the GSM$^2$ model. Specifically, we consider the three usual biological pathways, proposed in \cite{cordoni2021generalized}, which are a repair pathway stating that each sub-lethal lesion is repaired at rate $r$, a death pathway of sub-lethal lesions $X$ to become a lethal one $Y$ at rate $a$, and another death pathway for the pairwise interaction of two sub-lethal lesions $X$ at rate $b$. These events occur according to the following Markovian biological processes,
\begin{equation}\label{EQ:React}
\begin{cases}
X^{(j,d)} \xrightarrow{r} \emptyset^{(j,d)}\,,\\
X^{(j,d)} \xrightarrow{a} Y^{(j,d)}\,,\\
X^{(j,d)} + X^{(j,d)} \xrightarrow{b} Y^{(j,d)}\,,
\end{cases}
\end{equation}
as summarized in Figure \ref{fig:GSM2_scheme_NTCP} (iii). Finally, the GSM$^2$ model predicts the survival probability $S_j$ of each cell $j$ \cite{cordoni2022cell, battestini2024multiscale}, according to the following equation, 
\begin{equation}\label{EQ:surv}
S_j:=\mathbb{P}\left ( \lim_{t \to \infty} \left(Y^{(j,1)}(t),\dots,Y^{(j,N_d)}(t)\right) = \mathbf{0} \right)\,,
\end{equation}
which is the input of the organ response model. It has been shown in \cite{cordoni2022cell} that $S_j$ can be computed analytically as
\begin{equation}\label{EQN:S_GSM2}
S_j = 
\begin{cases}
\prod_{d=1}^{N_d}\frac{rx_{j,d}}{(a+r)x_{j,d} + bx_{j,d}(x_{j,d}-1)}\, & \mbox{if} \qquad y_{j,d} = 0\,,\qquad \forall d = 1\,,\dots, N_d\,,\\
0 & \mbox{otherwise}\,,
\end{cases}
\end{equation}
being $x_{j,d}$ and $y_{j,d}$, respectively, the initial number of sub-lethal and lethal lesions, respectively, for domain $d = 1\,,\dots, N_d$, as described in Sections \ref{Subsubsection: physics}-\ref{Subsubsection: DNA damage}.

In the case of fractionated irradiation, namely when we deliver a total prescribed dose $D^{\mathrm{tot}}$ in a number of fractions $N^{\mathrm{frac}}$, which means a dose per fraction of $D^{\mathrm{frac}}= D^{\mathrm{tot}} / N^{\mathrm{frac}}$, the complete cell survival probability is computed as
\begin{equation}\label{EQ:surv_Nfrac}
S_j^{\mathrm{tot}} = \left( S_j \right) ^{N^{\mathrm{frac}}}\,.
\end{equation}
 
\subsection{The whole organ/tumor effect model}
\label{Subsection: organ response}

We model the radiation-induced organ and tumor response, as illustrated in Figure \ref{fig:GSM2_scheme_NTCP} (iv), by integrating the single-cell GSM$^2$ model into the RSM \cite{kallman1992tumour}, thus developing a general mechanistic-driven serial/parallel organ model. In this way, the model can consider volume effects, namely, how an organ responds to damage depending on its tissue architecture and irradiated volume. It is worth stressing that the single‑cell resolution of the model, combined with the stochastic nature of microdosimetry, enables the modeling of fully heterogeneous radiation fields, both in terms of dose and radiation quality. In addition, the proposed model can capture heterogeneous environments of the irradiated organs, considering radiation-mediated stochasticities in the induction and evolution of DNA damages. 

We assume that an organ comprises FSUs constituting the organ's basic functioning units. Thus, a single FSU is the minimum ensemble of clonogens that preserves organ functionality. Different organs have different arrangements and sizes of FSUs. \cite{bidanta2023functional} reports the typical size considered for different organs relevant to radiotherapy. In this study, for simplicity, we model the FSUs as cubic boxes with a given side length $l_{FSU}$. However, it is possible to describe FSUs of any geometric shape and size. 

As standard, \cite{niemierko1993modeling}, we assume that a single surviving clonogen in an FSU is enough for the $i$-th FSU to repopulate and survive. We define $P_i^{\mathrm{FSU}}(D_i, \mathrm{LET}_i, [\mathrm{O_2}]_i)$ as the probability that all clonogens in the FSU die given an average macroscopic dose $D$ and LET, also considering an oxygenation level $[\mathrm{O_2}]$. Modeling the FSU as outlined, we can calculate its biological response to IR as
\begin{equation} \label{EQ:FSU_die}
P_i^{\mathrm{FSU}}(D_i, \mathrm{LET_i}, [\mathrm{O_2}]_i) := \prod_{j=1}^{N^{cell}_i} \left(1 - S_j\left(z_n^j, \mathrm{LET}^j, X^j, Y^j, [\mathrm{O_2}]^j |D_i, \mathrm{LET}_i\right)\right),
\end{equation}
where the index $j$ spans all clonogens $N^{cell}_i$ inside the $i$-th FSU.

Also, $S_j(z_n^j, \mathrm{LET}^j, X^j, Y^j, [\mathrm{O_2}]^j | D_i, \mathrm{LET}_i)$ represents the survival probability of the $j$-th clonogen, according to equation \eqref{EQ:surv}, after receiving a local dose $z_n^j$ and radiation quality $\mathrm{LET}^j$, considering an average macroscopic dose $D$ and LET, along with a distribution of sub-lethal and lethal lesions $X^j$ and $Y^j$, respectively. A notable distinction from most existing models is that each clonogen possesses a unique survival probability due to the stochastic nature of energy deposition and resulting damage induction. Moreover, all clonogens are correlated through the average macroscopic dose $D$ and LET. Lastly, we specify that in the case of fractionated irradiation, the single-cell surviving probability in equation \eqref{EQ:FSU_die} is computed according to equation \eqref{EQ:surv_Nfrac}.

The organ response to radiation is thus determined by the organ structure, described by how these FSUs are geometrically and/or functionally interconnected. In particular, it is defined as a fully parallel organ that requires all of its FSUs to die to be inactivated. On the contrary, a fully serial organ is inactivated as soon as a single FSU is inactivated, as previously discussed. Figure \ref{fig:organs_FSU} reports a depiction of different organ structures: (a) a fully serial, (b) a fully parallel, and (c) a complex mixed organ structure.

The response, in the sense of organ inactivation, of a completely serial organ composed of $m$ FSUs, e.g., Figure \ref{fig:organs_FSU} (a), each of which received a possibly different macroscopic average dose $D_i$ and LET$_i$, $i = 1,\dots, m$, is computed as the probability of at least one FSU to die, that is
\begin{equation}\label{EQN:Serial}
P^{\mathrm{serial}} = 1 - \prod_{i=1}^m \left(1 - P_i^{\mathrm{FSU}}\left(D_i, \mathrm{LET}_i, [\mathrm{O_2}]_i\right)\right)\,.
\end{equation}
Thus, inactivating any FSU leads to complications. In this case, we typically refer to it as a \textit{small volume effect}.

On the contrary, the inactivation due to radiation of a fully parallel organ with $n$ FSUs, as reported in Figure \ref{fig:organs_FSU} (b), is calculated as the probability of all FSUs to die, that is
\begin{equation}\label{EQN:Parallel}
P^{\mathrm{parallel}} = \prod_{i=1}^n P_i^{\mathrm{FSU}}\left(D_i, \mathrm{LET}_i, [\mathrm{O_2}]_i\right)\,.
\end{equation}
So, any FSU remaining active preserves organ function. For this reason, we usually talk about a \textit{large volume effect}.

In addition, the intermediate situation can be considered as the inactivation of an organ composed of $m$ serial FSUs and $n$ parallel FSUs, e.g., Figure \ref{fig:organs_FSU} (c). So we define the NTCP as
\begin{equation}\label{EQN:SerialParallel}
\mathrm{NTCP} = \prod_{k=1}^n \left[ 1 - \prod_{i=1}^m \left(1 - P_{i,k}\left(D_{ik}, \mathrm{LET}_{ik}, [\mathrm{O_2}]_{ik}\right)\right) \right ]\,.
\end{equation}
which is a generalization of the RSM model \cite{kallman1992tumour}.

Since the typical situation in a clinical setting is to completely eradicate the tumor, the TCP is defined as the probability of killing all the cells in the tumor. We thus obtain
\begin{equation}\label{EQN:TCP}
\mathrm{TCP} = \prod_{j=1}^{N^{cell}} \left(1 - S_j\left(z_n^j, \mathrm{LET}^j, [\mathrm{O_2}]^j, X^j, Y^j |D, \mathrm{LET}\right)\right)\,,
\end{equation}
taking into account the characteristic oxygen distribution $[\mathrm{O_2}]^j$ of the cancer volume with a cell level resolution.

Regarding normal tissue, in analogy to \cite{kallman1992tumour}, when considering homogeneous radiation quality and dose distribution, uniform oxygenation level in the whole target volume, and, lastly, identical sensitivity of all FSUs, which means $P_{i,k} \equiv P^{\mathrm{FSU}}, \ \forall i,j$. Thus, equation \eqref{EQN:SerialParallel} becomes:
\begin{equation}\label{EQN:SerialParallel_homo}
\mathrm{NTCP} = \left[ 1 - \left[1 - P^{\mathrm{FSU}}\left(D, \mathrm{LET}, [\mathrm{O_2}]\right)\right]^m \right ]^n\,.
\end{equation}

\begin{comment}

\marco{By inverting equation \eqref{EQN:SerialParallel_homo}, we obtain:}
\begin{equation}\label{EQN:FSU_homo}
\marco{P^{\mathrm{FSU}}\left(D, \mathrm{LET}, [\mathrm{O_2}]\right) = 1 - \left(1 - \mathrm{NTCP}^{1/n} \right)^{1/m}\,.}
\end{equation}

\marco{Now we want to estimate the probability of inducing a complication $P_{ab}$ only for a fraction $a \cdot b$ of the whole target volume, due to partial irradiation, where $0 \leq a \leq 1$ and $0 \leq b \leq 1$ are the relative fractions of the parallel and serial FSUs, respectively. Thus, considering Equations \ref{EQN:SerialParallel_homo} and \ref{EQN:FSU_homo}, we obtain:}
\begin{equation}\label{EQN:SerialParallel_homo_ab}
\marco{P_{ab} = \left[ 1 - \left(1 - P^{\mathrm{FSU}}\left(D, \mathrm{LET}, [\mathrm{O_2}]\right)\right)^{a \cdot m} \right ]^{b \cdot n} = \left[ 1 - \left(1 - \mathrm{NTCP}^{1/n}\right)^{a} \right ]^{b \cdot n} \,.}
\end{equation}

\end{comment}

Since complete knowledge of an organ structure is not always possible, the above organ response can be parameterized in terms of a seriality parameter $s$, which is defined as the number of serial FSUs, $m$, with respect to the total number of FSUs, which is $M = n \cdot m$, i.e.
\begin{equation} \label{EQN:s}
    s := \frac{m}{n \cdot m} = \frac{1}{n} \,,
\end{equation}
with the assumption that all parallel FSUs have been irradiated, according to \cite{kallman1992tumour}. In particular, $s = 1$ corresponds to a fully serial organ, while $s \rightarrow 0$ describes a fully parallel organ.

Then, by expressing 
\begin{equation}\label{EQN:EqAssunz}
    \left(1 - P^{\mathrm{FSU}}\right)^m = \left( \left(1 - P^{\mathrm{FSU}}\right)^s \right)^{m/s} = \left( \left(1 - P^{\mathrm{FSU}}\right)^s \right)^{m \cdot n} =  \left( \left(1 - P^{\mathrm{FSU}}\right)^s \right)^{M}\,,
\end{equation}
and substituting equation \eqref{EQN:EqAssunz} in equation \eqref{EQN:SerialParallel_homo}, we obtain:
\begin{equation}\label{EQN:SerialParallel_homo_s}
\mathrm{NTCP} = \left[ 1 - \left( \left[1 - P^{\mathrm{FSU}}\left(D, \mathrm{LET}, [\mathrm{O_2}]\right)\right]^s \right)^{M} \right ]^{^{\frac{1}{s}}} \,.
\end{equation}

\begin{comment}

\marco{From equation \eqref{EQN:SerialParallel_homo_ab}, we compute the response of a fraction of the whole volume $v = a \cdot 1$, using the seriality parameter of equation \eqref{EQN:s}, as}
\begin{equation}\label{EQN:SerialParallel_homo_ab_s}
\marco{P_{v} = \left[ 1 - \left(1 - \mathrm{NTCP}^{s}\right)^{v} \right ]^{1/s} \,.}
\end{equation}

\marco{Then, considering $a=1/m$, i.e. $v=1/m$, we have $P_{v} = P^{\mathrm{FSU}}$. Thus, equation \eqref{EQN:SerialParallel_homo_ab_s} becomes}
\begin{equation}\label{EQN:SerialParallel_homo_FSU_s}
\marco{P^{\mathrm{FSU}} = \left[ 1 - \left(1 - \mathrm{NTCP}^{s}\right)^{1/m} \right ]^{1/s} \,.}
\end{equation}
  
\marco{By inverting the last equation, we obtain:}
\begin{equation}\label{EQN:SerialParallel_homo_ab_s}
\marco{\mathrm{NTCP} = \left[ 1 - \left(1 - \mathrm{NTCP}^{s}\right)^{v} \right ]^{1/s} \,.}
\end{equation}

\marco{Now, by substituting equation \eqref{EQN:s} in equation \eqref{EQN:SerialParallel_homo}, we obtain:}
\begin{equation}\label{EQN:SerialParallel_homo_s}
\marco{\mathrm{NTCP} = \left[ 1 - \left(1 - P^{\mathrm{FSU}}\left(D, \mathrm{LET}, [\mathrm{O_2}]\right)\right)^m \right ]^n\,.}
\end{equation}

\end{comment}

Lastly, by generalizing equation \eqref{EQN:SerialParallel_homo_s} for heterogeneous radiation quality, dose, and oxygen distributions, i.e., $P_{i} \neq P_{i'}$, the NTCP of equation \eqref{EQN:SerialParallel} can also be computed as
\begin{equation}\label{EQN:NTCP}
\mathrm{NTCP} = \left[ 1 - \prod_{i=1}^M \left[1 - P_{i}\left(D_{i}, \mathrm{LET}_{i}, [\mathrm{O_2}]_i\right]^s \right) \right ]^{\frac{1}{s}}\,,
\end{equation}
which is the corrected version of the expression proposed in \cite{kallman1992tumour}.

Equations \eqref{EQN:Serial}, \eqref{EQN:Parallel}, \eqref{EQN:SerialParallel}, \eqref{EQN:TCP}, and \eqref{EQN:NTCP} mark a substantial difference with existing models, where the FSU inactivation $P_{i} \left(D_{i}, \mathrm{LET_{i}}, [\mathrm{O_2}]_i \right)$ is often calculated assuming Poisson statistics, averaging over the entire cell population, as in \cite{kallman1992tumour, Cometto2014}. 

In summary, the GSM$^2$‑driven NTCP and TCP modeling framework enables the investigation of a wide range of experimental scenarios. In particular, it can account for multiple parameters and biological effects, spanning from single‑cell resolution up to whole‑organ responses, including, for example:
\begin{enumerate}
    \item radiation quality (particle type, energy, and $\mathrm{LET}$);
    \item dose prescription;
    \item fractionation scheme (number of fraction, dose per fraction, and total dose);
    \item tissue heterogeneity, both in the sense of different environmental oxygenation levels and cell line radiosensitivity;
    \item tissue architecture and volume effect.
\end{enumerate}
Furthermore, physically, the model can account for energy deposition heterogeneity and stochasticity, while biologically, it describes multiple effects at different scales, ranging from well-known ones at the cellular level, such as the \textit{oxygen fixation effect} on DNA damage formation, to complex whole-organ responses. Lastly, thanks to the single-cell resolution provided by the GSM$^2$ model, the proposed mechanistic-driven NTCP and TCP models can effectively manage the effects of physical and tissue environmental heterogeneities, such as spatial dose, oxygenation levels, and cell line heterogeneity. 

\subsection{Numerical implementations}
\label{Subsection: num impl}

We show the stepwise construction and the computational information of the GSM$^2$-driven NTCP and TCP models, reported in equations \eqref{EQN:SerialParallel}, \eqref{EQN:NTCP}, and \eqref{EQN:TCP}. To do this, we performed the following steps:

\begin{enumerate}
    \item We consider an organ, divided by $m$ by $n$ cubic boxes of side length $l_{FSU}$, i.e. $M = m \cdot n$ FSUs, that represent the structural arrangement of the tissue. Each of these FSUs contains $N^{cell}_i$ cells, i.e., a cell population, as previously described. 
    \begin{enumerate}
        \item The healthy organ is characterized by a volume effect parameter $s$, according to equation \eqref{EQN:s}, and a distribution of oxygen concentration in which each cell $j$ is characterized by an oxygenation level $[\mathrm{O_2}]^j$, without compromising computational performance. In this part of the work, since we consider healthy organs, where oxygen variability is less relevant than in tumor tissues, we assume a uniform oxygenation $[\mathrm{O_2}]$ for simplicity. 
        \item Instead, we model the tumor volume as a sphere of cancer cells, mimicking the biological structure of a spheroid. In addition, we consider an oxygenation gradient in the spheroid, which means that each cell $j$ has a specific oxygenation level $[\mathrm{O_2}]^j$. Lastly, for the spheroid, we consider co-cultures, namely cells characterized by a different radiosensitivity, which means that each cell $j$ has a specific set of biological rates, i.e. $a^j$, $b^j$, and $r^j$.
    \end{enumerate}
    \item Given the radiation quality of the primary radiation, the radius of the beam $R_{beam}$, and a macroscopic prescribed dose $D$, we calculate the impact point of all the track hits.
    \item We calculate the radial dose distribution on the nucleus of each cell, and according to the AT model in equation \eqref{EQ:ATdose}, we obtain the absorbed dose $z_d$ for each particle $\nu$ in each domain $d$.
    \item We compute the initial radiation-induced sub-lethal and lethal lesions, i.e. $X^{(j,d)}$ and $Y^{(j,d)}$, respectively, according to equation \eqref{eq:kDNA}, for each domain $d$ of each cell $j$.
    %\item We simulate the evolution pathways via a stochastic simulation algorithm (SSA), given the initial number of lesions, according to the biological pathways in equation \eqref{EQ:React};
    \item We calculate the survival probability of each cell from the lesion's distributions according to Equation \eqref{EQN:S_GSM2}.
    \item For healthy tissue only, we compute the probability that all clonogens in the FSU die according to equation \eqref{EQ:FSU_die}, repeating the procedure for each FSU.
    \item We predict the NTCP or the TCP according to equations \eqref{EQN:SerialParallel}-\eqref{EQN:NTCP} or \eqref{EQN:TCP}, respectively, for the irradiated organ or tumor, respectively.
\end{enumerate}
The previous physical processes cannot be computed independently for each cell or FSU because they simultaneously involve more cells or FSUs. Instead, the computation of each FSU response can be performed independently. 

From a computational point of view, the GSM$^2$-driven dose-response models were implemented using \textit{Julia} \cite{bezanson2017julia} following the above stepwise construction. All the simulations was performed using an \textit{Intel(R) Core(TM) i9-14900, 2.00 GHz, RAM 64.0 GB, 31 cores}.

\subsection{Target volumes, irradiation parameters, biological endpoints, and radiobiological indices}
\label{Subsection: indeces}

In the case of healthy tissues investigations, we perform simulations for different radiation qualities of the projectile, characterized by particle type, energy, and LET. In particular, we focus on $100~\mathrm{MeV}$ protons, $150~\mathrm{MeV/u}$ and $80~\mathrm{MeV/u}$ helium ions, and $280~\mathrm{MeV/u}$, $140~\mathrm{MeV/u}$, and $80~\mathrm{MeV/u}$ carbon ions, with a beam radius $R_{beam} = 5~\mathrm{mm}$. We consider a parallelepiped which mimics an organ of volume $3.9~\mathrm{mm}^3$, divided by $m$ by $n$ cubic boxes of side $300~\mathrm{\mu m}$, i.e., $M = m \cdot n$ FSUs, that represent the structural arrangement of the tissue. Each FSU contains $N^{cell}_{i}$ cells of radius $15~\mathrm{~\mu m}$, i.e., a cell population with a nucleus radius of $7.2~\mathrm{\mu m}$ and a domain radius of $0.8~\mathrm{\mu m}$. For this \textit{in-silico} study, we consider realistic values for the three biological rates of the GSM$^2$, namely $a = 0.01$, $b = 0.3$, and $r = 4.3$, which are within typical ranges as discussed in \cite{battestini2024multiscale}. For a fixed value of the organ and FSU volumes, we study the impact of different arrangement of the $M$ FSUs, i.e. the different tissue seriality according to equation \eqref{EQN:s}, from a mostly parallel ($s=0.02$) to a completely serial ($s=1.0$) situation, on the biological outcome for the whole organ, in the case of different doses per fraction and fractionation schemes. We also study the role of the partial irradiation (PI) of the organ, delivering a uniform dose only on a fraction of the whole organ ($\mathrm{PI} = 1/3, 2/3$), repeating the procedure for different dose levels, tissue seriality parameters, radiation qualities (particle type, energy and $\mathrm{LET}$), and environmental oxygen concentrations, on the emergence of a complication for that organ. In particular, we consider different constant oxygenation levels, from a hypoxic ($0.5\%$ $\mathrm{O_2}$) to a physioxic condition ($7\%$ $\mathrm{O_2}$).

Instead, for tumor analyses, we consider a sphere of volume $\sim 0.3~\mathrm{mm}^3$, filled with cells at different oxygenation levels, from a hypoxic inner core ($0.1\%$ $\mathrm{O_2}$), with a radius of $\sim 50~\mathrm{\mu m}$, to a physioxic outer layer ($7\%$ $\mathrm{O_2}$). This oxygen distribution is compared to a uniform mean oxygenation ($ \sim 5\%$ $\mathrm{O_2}$), calculated as the average oxygenation across all spheroid cells in the gradient setup. The dimensions considered for the clonogenes of the irradiated volumes are in agreement with the typical cell density of $10^4-10^5$ cells$/$mm$^3$ \cite{kallman1992tumour}. We irradiated it using a $100~\mathrm{MeV}$ proton beam and a $80~\mathrm{MeV/u}$ carbon ion beam. We consider $a = 0.01$, $b = 0.3$, and $r = 4.3$ for this simulation. In addition, for the second part of this investigation, we consider a spheroid with the same volume as the previous one, characterized by a uniform oxygenation of $ 7\%$ $\mathrm{O_2}$, but made by two cell populations with different radiosensitivity. In detail, $42\%$ or $10\%$ of the spheroid is composed of \textit{radio-resistant} cells, characterized by the parameters $a = 0.0059$, $b = 0.058$, and $r = 5.84$, while the remaining $58\%$ or $90\%$, respectively, consists of \textit{radio-sensitive} cells, with $a = 0.013$, $b = 0.040$, and $r = 2.78$. These parameter sets reproduce the survival fractions reported in~\cite{sinclair1968cyclic} for different levels of radiosensitivity and correspond, for the reference radiation, to $\alpha = 0.1235$ and $\beta = 0.0285$ for the resistant subpopulation, and $\alpha = 0.793$ and $\beta = 0$ for the sensitive one. The radiosensitivity level is randomly distributed in the spheroid's cells. This co-culture is compared with spheroids composed of only resistant or sensitive cells.

To compare the various NTCP and TCP curves obtained after simulations for different physical irradiation parameters and biochemical environmental conditions, we compute the whole organ dose corresponding to NTCP/TCP of 50$\%$, i.e., $\mathrm{TD_{50}}$. 

Furthermore, since the shape of the NTCP curve is influenced not only by the value of $\mathrm{TD_{50}}$ but also by its steepness around this dose level, we introduce a parameter that quantifies the mid-range slope of the NTCP curve. We denote this quantity as the $\mathrm{TD_{25-75}}$\emph{-slope} ($\mathrm{TS}_{25-75}$), defined as
\begin{equation} \label{eq:TS50}
    \mathrm{TS}_{25-75} = \frac{0.75 - 0.25}{\mathrm{TD_{75}} - \mathrm{TD_{25}}} = \frac{0.5}{\mathrm{TD_{75}} - \mathrm{TD_{25}}} \, .
\end{equation}
This parameter represents the chord slope of the NTCP curve between the points $(\mathrm{TD_{25}},\,0.25)$ and $(\mathrm{TD_{75}},\,0.75)$, thus providing an intrinsic measure of the curve steepness in the vicinity of $\mathrm{TD_{50}}$, which is typically called $m_{50}$ in literature. $\mathrm{TD_{50}}$ and $m_{50}$ are the typical indices that characterize the dose-response relationship in RT. 

Lastly, to compute the shift between NTCP or TCP curves at different conditions, which quantifies the biological effectiveness of PT at a given scenario, we define a dose modifying factor (DMF) as the ratio between the $\mathrm{TD_{50}}$ at the reference condition ($\mathrm{c_r}$) and the $\mathrm{TD_{50}}$ at the $x$-th irradiation and environmental condition ($\mathrm{c_x}$) to achieve the same level of biological effect, i.e. NTCP or TCP of 50$\%$, that is
\begin{equation} \label{EQN:DMF}
    \mathrm{DMF_{50}} := \left . \frac{\left( \mathrm{TD_{50}} \right) ^{\mathrm{c_r}}}{\left(\mathrm{TD_{50}} \right)^{\mathrm{c_x}}} \right |_{\mathrm{NTCP, TCP = 0.5}}\,.
\end{equation}

%% RESULTS
\section{Results}
\label{2_Results}
\counterwithin{figure}{section}
\setcounter{figure}{0}

In the following sections, we show the numerical results obtained through the GSM$^2$-driven NTCP and TCP models.

\begin{figure}[htbp]
\centering
\centering
\includegraphics[width=0.8\linewidth]{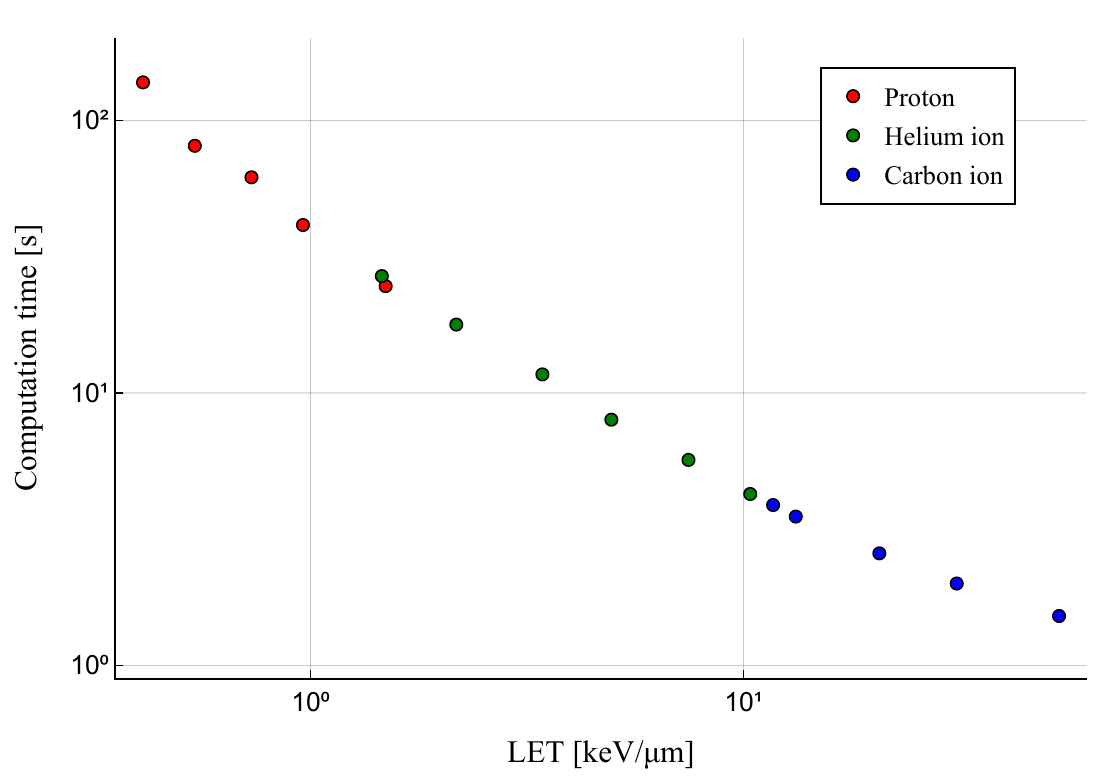}
\caption{Computational times for 1 Gy delivered on a box of $1~\mathrm{mm}^3$ that contains 35937 cells. Irradiation beam of radius $0.7~\mathrm{mm}$ for different radiation qualities, in particular proton (red points), helium ion (green points), and carbon ion (blue points) beams with various LET values.}\label{FIG:Compt_time} 
\end{figure}

Figure \ref{FIG:Compt_time} reports the computation times for 1 Gy irradiation of a box of $1~\mathrm{mm}^3$ that contains 35937 cells with a beam of radius $R_{beam} = 0.7~\mathrm{mm}$ for different radiation qualities, in particular proton, helium ion, and carbon ion beams with various LET values.

\subsection{NTCP predictions for different tissue environments and irradiation conditions}
\label{Subsubsection: NTCP predictions}

Here, we focus on studying the impact of several physical and biological parameters on modulating the dose-effect relationship. In particular, we investigate the feasibility of the GSM$^2$-driven NTCP model in describing various experimental scenarios across different beam radiation qualities. We will focus on the results obtained for two representative radiation-quality cases, i.e., $100~\mathrm{MeV}$ protons and $280~\mathrm{MeV/u}$ carbon ions, typically used in PT. We analyze the impact of different macroscopic doses, fractionation schemes, and partial irradiations on the emergence of side effects. In addition, we consider different levels of environmental oxygen and various organs with specific volume effects, as described in the previous section.

\begin{figure}[htbp]
\centering
\centering
\includegraphics[width=0.9\linewidth]{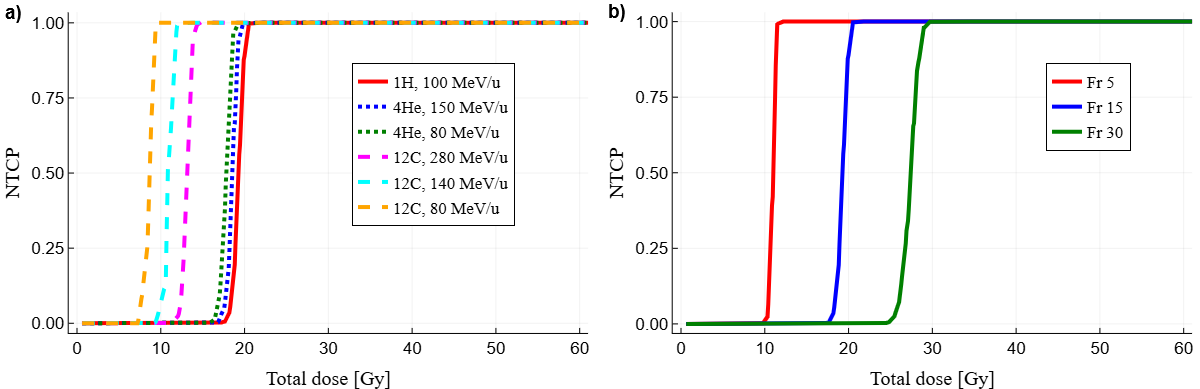}
\caption{Irradiation of a healthy organ ($s=1.0$ and $7\%$ $\mathrm{O_2}$). \textbf{(a)} NTCP as a function of total dose, across different LET values, using proton (solid line), helium ion (dotted line), and carbon ion (dashed line) beams, with 15 fractions. \textbf{(b)} NTCP as a function of total dose, across different fractionation schemes, i.e., 5 (red), 15 (blue), and 30 (green) fractions for the same total dose, with $100~\mathrm{MeV}$ proton beam.}\label{fig:NTCP_LET_Fr} 
\end{figure}

Figure \ref{fig:NTCP_LET_Fr} reports the NTCP curves as a function of the total dose of a given organ at risk ($s=1.0$) at a uniform oxygenation level of $7\%$ $\mathrm{O_2}$, for different irradiation conditions. In detail, Figure \ref{fig:NTCP_LET_Fr}~(a) shows the influence of different LET values, using $100~\mathrm{MeV}$ proton, $150~\mathrm{MeV/u}$ and $80~\mathrm{MeV/u}$ helium ion, and $280~\mathrm{MeV/u}$, $140~\mathrm{MeV/u}$, and $80~\mathrm{MeV/u}$ carbon ion beams, and 15 fractions, while Figure \ref{fig:NTCP_LET_Fr}~(b) highlights the impact of three different fractionation schemes, namely 5, 15, and 30 fractions for the same total dose, in the case of a $100~\mathrm{MeV}$ proton beam.

\begin{figure}[htbp]
\centering
\centering
\includegraphics[width=0.9\linewidth]{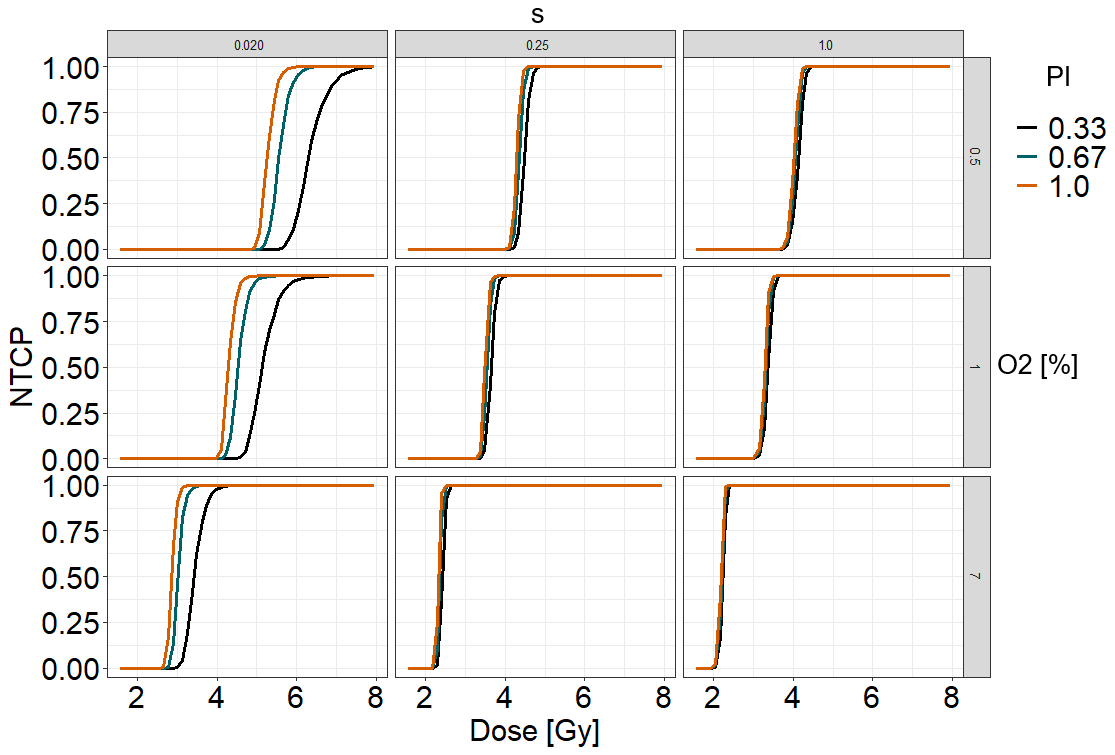}
\caption{NTCP as a function of dose per fraction in the case of $100~\mathrm{MeV}$ proton beam irradiation in 5 fractions. The horizontal panels report NTCP curves for different seriality parameters ($s=0.02, 0.25, 1.0$), while the vertical panels show NTCP curves for various environmental oxygenation levels ($0.5\%, 1.0\%, 7.0\%$ $\mathrm{O_2}$). Different colors correspond to different partial irradiation of the organ volume ($\mathrm{PI} =$ 0.33 (black), 0.67 (green), 1.0 (orange)).}\label{fig:NTCP_p} 
\end{figure}

\begin{figure}[htbp]
\centering
\centering
\includegraphics[width=0.9\linewidth]{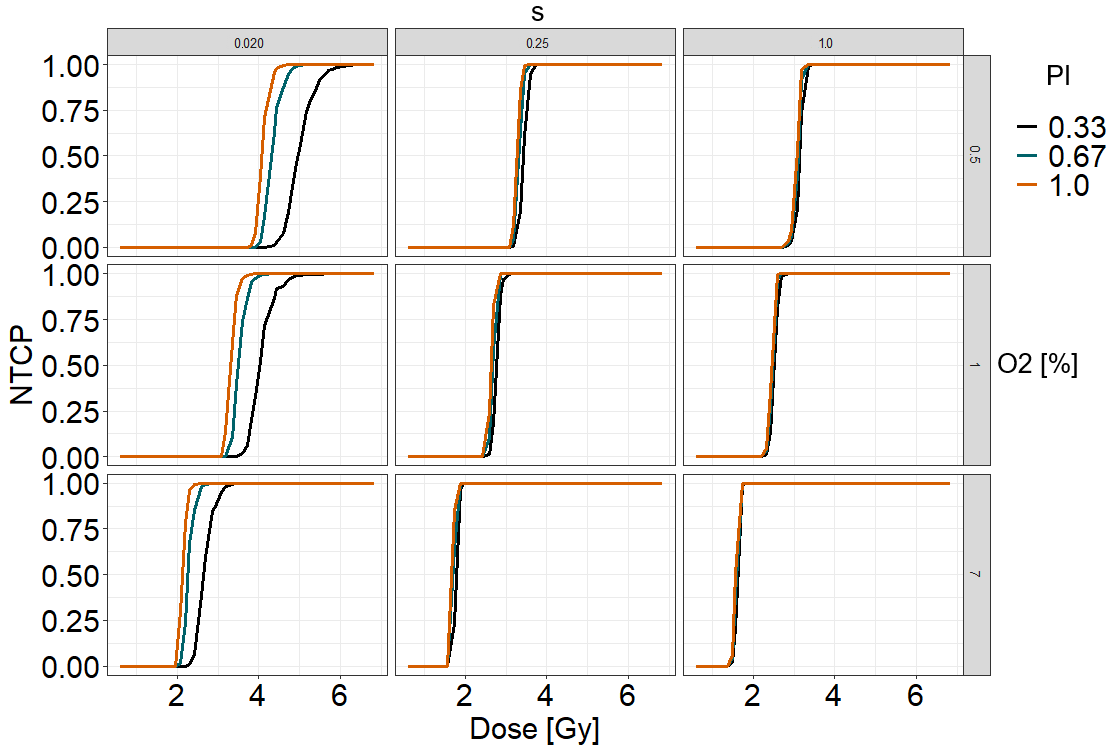}
\caption{NTCP as a function of dose per fraction in the case of $280~\mathrm{MeV/u}$ carbon ion beam irradiation in 5 fractions. The horizontal panels report NTCP curves for different seriality parameters ($s=0.02, 0.25, 1.0$), while the vertical panels show NTCP curves for various environmental oxygenation levels ($0.5\%, 1.0\%, 7.0\%$ $\mathrm{O_2}$). Different colors correspond to different partial irradiation of the organ volume ($\mathrm{PI} =$ 0.33 (black), 0.67 (green), 1.0 (orange)).}\label{fig:NTCP_C} 
\end{figure}

Figures \ref{fig:NTCP_p}-\ref{fig:NTCP_C} display the influence of environmental oxygen concentration ($0.5\%, 1.0\%, 7.0\%$ $\mathrm{O_2}$), tissue seriality ($s=0.02, 0.25, 1.0$), and partial irradiation ($\mathrm{PI} = 0.33, 0.67, 1.0$) of a given organ at risk, on NTCP as a function of the dose per fraction, for two different radiation particles and energies. NTCP curves for both particles are obtained from equation \eqref{EQN:NTCP}. In detail, Figure \ref{fig:NTCP_p} shows the NTCP curves in the case of $100~\mathrm{MeV}$ proton beam irradiation, while Figure \ref{fig:NTCP_C} reports the NTCP curves in the case of $280~\mathrm{MeV/u}$ carbon ion beam irradiation, for an irradiation scheme of 5 fractions.

\begin{figure}[htbp]
\centering
\centering
\includegraphics[width=0.9\linewidth]{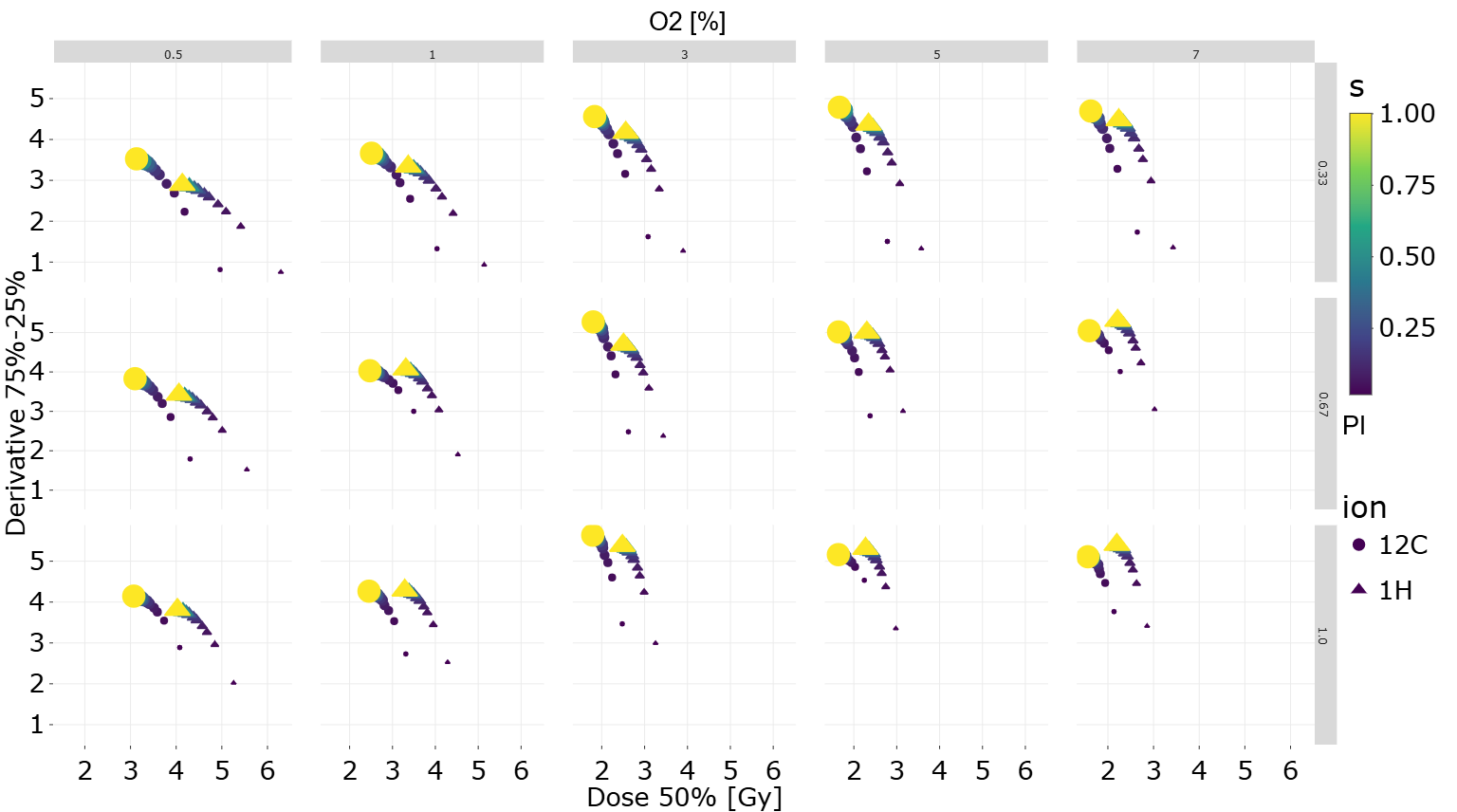}
\caption{$\mathrm{TD_{50}}$ (horizontal axes) and $\mathrm{TS}_{25-75}$ as a function of tissue seriality, i.e. $s \in (0,1]$ (color map and symbol dimension), for $100~\mathrm{MeV}$ proton beam (triangles) and $280~\mathrm{MeV/u}$ carbon ion beam (circles) irradiation in 5 fractions. The horizontal panels display the impact of various oxygen concentrations ($0.5\%, 1.0\%, 3.0\%, 5.0\%, 7.0\%$ $\mathrm{O_2}$), while the vertical panels show the impact of different partial irradiation of the organ volume ($\mathrm{PI} = 0.33, 0.67, 1.0$).}\label{fig:NTCP_index_p} 
\end{figure}

Figure \ref{fig:NTCP_index_p} shows the whole organ dose (per fraction) corresponding to NTCP of 50$\%$, i.e. $\mathrm{TD_{50}}$, and the slope of the dose-response curve in proximity of $\mathrm{TD_{50}}$, defined as the derivative $\mathrm{TS}_{25-75}$ between the points $(\mathrm{TD_{25}},\,0.25)$ and $(\mathrm{TD_{75}},\,0.75)$ in equation \eqref{eq:TS50}, as a function of tissue seriality ($s \leq 1.0$), for $100~\mathrm{MeV}$ proton beam (triangles) and $280~\mathrm{MeV/u}$ carbon ion beam (circles) irradiation and 5 fractions, across different environmental oxygenation level ($0.5\%, 1.0\%, 3.0\%, 5.0\%, 7.0\%$ $\mathrm{O_2}$), and partial irradiation of the organ ($\mathrm{PI} = 0.33, 0.67, 1.0$).

%Validation on \textit{in-vivo} experiments

\subsection{TCP predictions for spheroids}
\label{Subsubsection: TCP predictions}

\begin{figure}[htbp]
\centering
\centering
\includegraphics[width=0.9\linewidth]{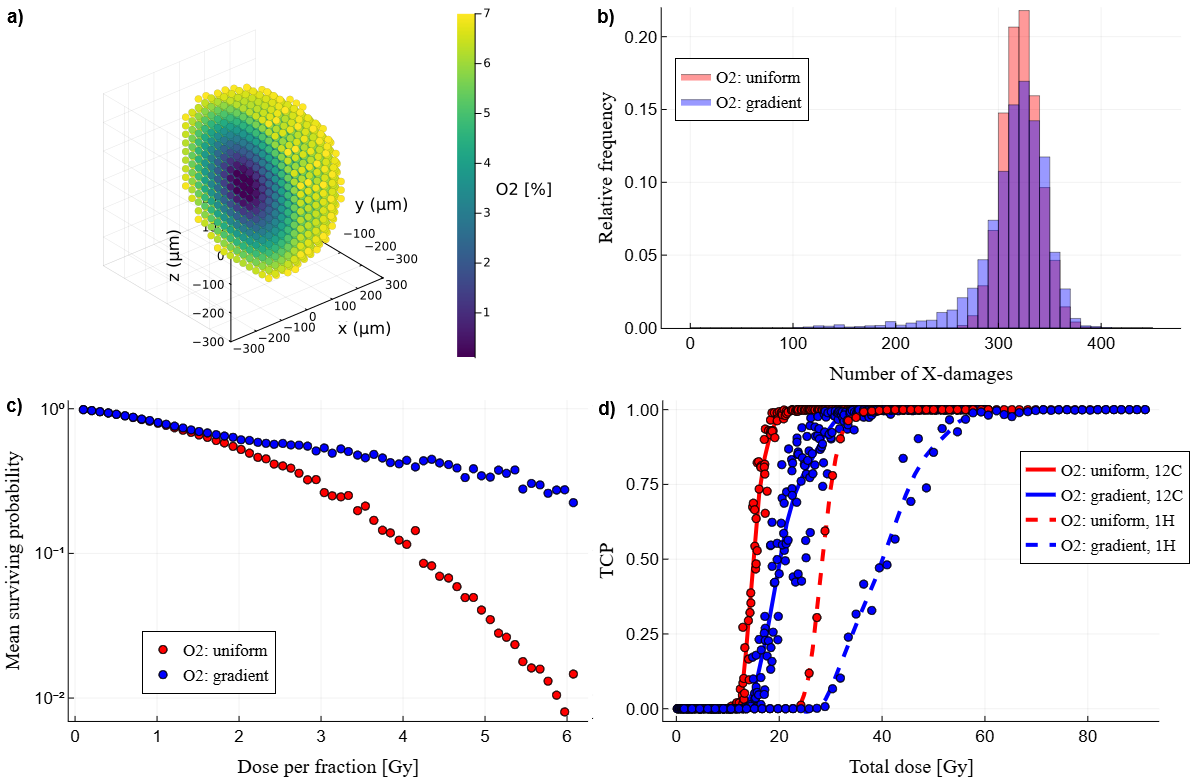}
\caption{Irradiation of a spheroid with a $100~\mathrm{MeV}$ proton beam and 15 fractions, considering hypoxic core of $0.1\%-7.0\%$ $\mathrm{O_2}$ and oxygenation gradient in the range $0.1\%-7.0\%$ $\mathrm{O_2}$ (blue lines and dots) or uniform mean oxygenation of $\sim5.0\%$ $\mathrm{O_2}$ (red lines and dots). \textbf{(a)} Spheroid geometry and oxygen gradient. Impact of oxygen distribution type in terms of \textbf{(b)} relative frequency of initial sub-lethal DNA damage for the spheroid’s cells in the case of 6 Gy, after 1 fraction, \textbf{(c)} average cell survival probability for all cells of the spheroid in a single fraction, i.e. $\sqrt[N^{\mathrm{frac}}]{\bar{S}^{\mathrm{tot}}}$, and \textbf{(d)} TCP as a function of total dose, comparing $100~\mathrm{MeV}$ proton beam and $80~\mathrm{MeV/u}$ carbon ion beam. The points are the results of the simulation, while the curves are obtained from a fitting procedure.}\label{fig:TCP_spheroid_O2} 
\end{figure}

Figure \ref{fig:TCP_spheroid_O2} shows the impact of different environmental oxygen distributions, i.e. an oxygenation gradient between $0.1 \%$ $\mathrm{O_2}$ (hypoxic inner core of radius $\sim 50~\mathrm{\mu m}$) and $7.0 \%$ $\mathrm{O_2}$ (physioxic outer layer), which is illustrated in Figure \ref{fig:TCP_spheroid_O2}~(a), compared to a uniform mean oxygenation level of $\sim 5.0 \%$ $\mathrm{O_2}$, in terms of different biological endpoints for a spheroid. In detail, Figure \ref{fig:TCP_spheroid_O2}~(d) displays TCP predictions, according to equation \eqref{EQN:TCP}, as a function of total dose, after spheroid irradiation with a $100~\mathrm{MeV}$ proton beam and a $80~\mathrm{MeV/u}$ carbon ion beam, in 15 fractions, for both environmental conditions, namely oxygenation gradient (blue line) and uniform mean oxygenation (red line). In detail, the points are the results of the GSM$^2$-driven TCP simulation, while the curves correspond to a loess regression fit of the data. The shift between the two TCP curves is quantified in terms of $\mathrm{DMF_{50}}$, according to equation \eqref{EQN:DMF}, and it is equal to 1.4 for protons and 1.3 for carbon ions, considering the gradient case as the reference condition. Instead, Figures \ref{fig:TCP_spheroid_O2}~(b)-(c) show the radiation-induced response at a cellular level for the $100~\mathrm{MeV}$ proton beam irradiation. In particular, \ref{fig:TCP_spheroid_O2}~(b) reports the initial sub-lethal DNA damage distributions, i.e. $X$ lesions according to equation \eqref{eq:kDNA}, after the deposition of 6 Gy in a single fraction, while \ref{fig:TCP_spheroid_O2}~(c) shows the mean cell survival probability in a single fraction, i.e. $\sqrt[N^{\mathrm{frac}}]{\bar{S}^{\mathrm{tot}}}$, which is obtained by averaging the final surviving probabilities of all cells, i.e. $\bar{S}^{\mathrm{tot}}$ from equation \eqref{EQ:surv_Nfrac}, and then by computing the $N^{\mathrm{frac}}$-th root, as a function of dose per fraction, for both oxygenation conditions.

\begin{figure}[htbp]
\centering
\centering
\includegraphics[width=0.9\linewidth]{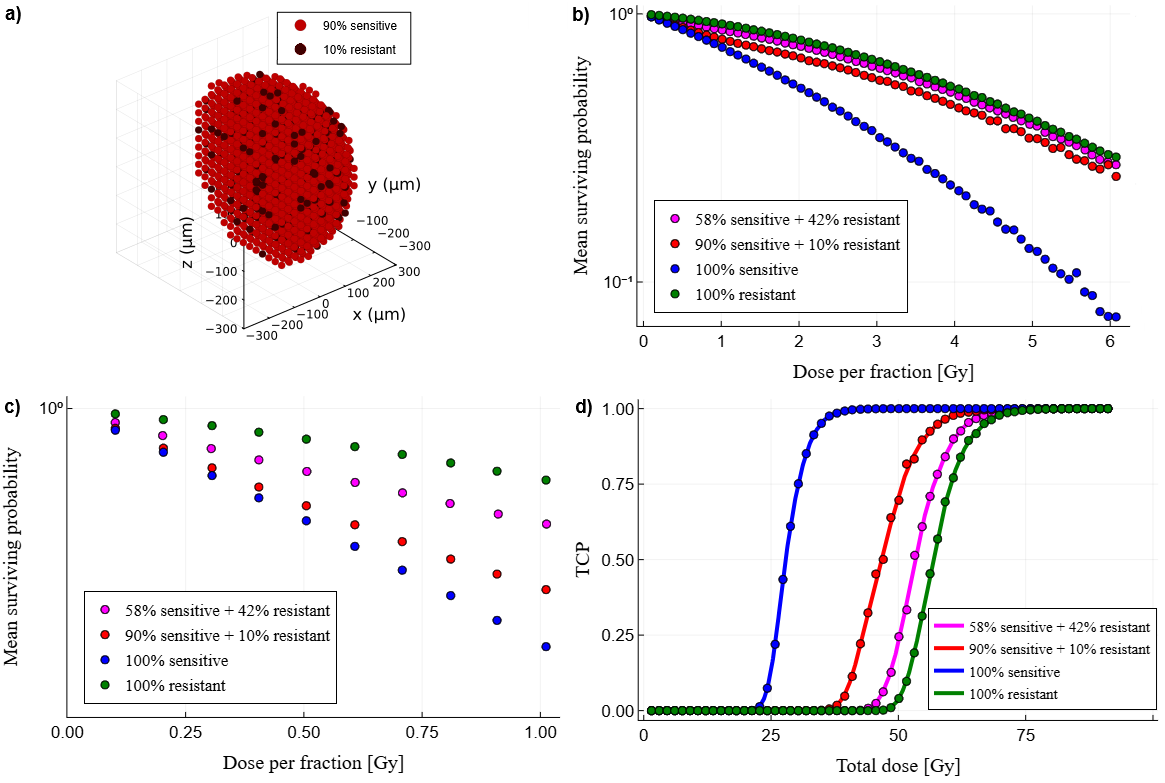}
\caption{Irradiation of a spheroid, at $7.0\%$ $\mathrm{O_2}$, with a $100~\mathrm{MeV}$ proton beam and 15 fractions, considering a radiosensitive cell line (blue lines and dots), a radioresistant cell line (green lines and dots), or a co-culture (magenta/red lines and dots), consisting of 58$\%$/90$\%$ sensitive line and 42$\%$/10$\%$ resistant line. \textbf{(a)} Spheroid geometry and cell lines distribution. Impact of cell type in terms of \textbf{(b)} average cell survival probability for all cells of the spheroid in a single fraction, i.e. $\sqrt[N^{\mathrm{frac}}]{\bar{S}^{\mathrm{tot}}}$, \textbf{(c)} low-dose zoom of the previous surviving probability, and \textbf{(d)} TCP as a function of total dose. The points are the results of the simulation, while the curves are obtained from a fitting procedure.}\label{fig:TCP_spheroid_co} 
\end{figure}

Figure \ref{fig:TCP_spheroid_co} shows the impact of the distribution of different cell lines, i.e., co-cultures that mimic the tumor microenvironment, consisting of two cell types (90$\%$ radiosensitive line and 10$\%$ radioresistant line, and 58$\%$-42$\%$), which is illustrated in Figure \ref{fig:TCP_spheroid_co}~(a), compared to a single radiosensitive cell line and a single radioresistant cell line, in terms of different biological endpoints for a spheroid at $7.0\%$ $\mathrm{O_2}$. In detail, Figure \ref{fig:TCP_spheroid_co}~(d) displays TCP predictions, according to equation \eqref{EQN:TCP}, as a function of total dose, after spheroid irradiation with a $100~\mathrm{MeV}$ proton beam in 15 fractions, for the four tissue conditions, namely co-culture (magenta and red lines), sensitive line (blue line), and resistant line (green line). In detail, the points are the results of the GSM$^2$-driven TCP simulation, while the curves correspond to a loess regression fit of the data. The shifts between the TCP curves, considering the resistant case as the reference condition, are quantified in terms of $\mathrm{DMF_{50}}$, according to equation \eqref{EQN:DMF}. These are equal to 1.1/1.2 for the radioresistant condition and the (58$\%$-42$\%$)/(90$\%$-10$\%$) co-culture, and to 2.0 for the radiosensitive and radioresistant conditions. Instead, Figures \ref{fig:TCP_spheroid_co}~(b-c) show the mean cell survival probability in a single fraction, i.e. $\sqrt[N^{\mathrm{frac}}]{\bar{S}^{\mathrm{tot}}}$, which is obtained by averaging the final surviving probabilities of all cells, i.e. $\bar{S}^{\mathrm{tot}}$ from equation \eqref{EQ:surv_Nfrac}, and then by computing the $N^{\mathrm{frac}}$-th root, as a function of dose per fraction, for three tissue conditions.

%\subsection{Validation on \textit{in-vivo} experiments}
%\label{Subsubsection: NTCP comparison}

%% DISCUSSION
\section{Discussion}
\label{3_Discussion}
\counterwithin{figure}{section}
\setcounter{figure}{0}

%Radiation-induced response of the rat spinal cord

The main goal of this work is to extend the GSM$^2$ from predicting \textit{in-vitro} preclinical data to developing a mechanistic-driven TCP/NTCP model, enabling a comprehensive investigation of clinically relevant \textit{in-vivo} endpoints for PT. This enables bridging the gap from the single-cell radiation-induced response to clinically relevant predictions.

Incorporating the GSM$^2$ model \cite{cordoni2021generalized, cordoni2022cell} into a mechanistic TCP/NTCP model enables the resulting tool to retain the core features of GSM$^2$. Specifically, it preserves the ability to account for different levels of stochasticity in physics and biology, particularly in energy deposition and the formation and progression of DNA damage, as discussed in \cite{cordoni2022multiple}. Further, as shown by our results, the stochastic behavior at the single‑cell level propagates to macroscopic tissue volumes, naturally incorporating tissue heterogeneity into the model. In particular, the main idea behind the development of the present TCP/NTCP model is to avoid an average description based on Poisson statistics and instead adopt a cell‑by‑cell representation. This is essential in PT, where average quantities cannot adequately capture the highly localized energy deposition characteristic of charged particles. By contrast, cell‑level stochasticity plays a crucial role in determining the final macroscopic biological effect \cite{Bellinzona2021}. Therefore, the present model goes beyond the Poisson assumption on the formation of initial DNA damage and the subsequent induction of complications, as discussed in \cite{cordoni2023emergence} for the original GSM$^2$ model, distinguishing this developed NTCP model from typical dose-response models, such as \cite{kallman1992tumour}. Incorporating a mechanistic radiobiological model into an established NTCP framework, such as the one adopted here from \cite{kallman1992tumour}, also facilitates future clinical translation. In particular, it enables the use of parameters like the seriality index $s$, which is extensively analyzed in this work.

The first technical aspect it is worth highlighting is the computational cost of the model. For the irradiation and geometrical conditions considered in this study, with a cell density range of $10^4-10^5$ per $\mathrm{mm^3}$, generating an NTCP or TCP curve in just a few seconds to minutes is possible. This conclusion is confirmed by the computational times for the simulation of energy deposition and dose distribution in the target volume reported in Figure \ref{FIG:Compt_time}, which is the most expensive part of the proposed model. In detail, the computation time decreases as LET increases, because the simulated number of particles needed to produce the same macroscopic dose in the target volume decreases, according to equation \eqref{EQ:Nparticle}.

\subsection{Impact of different irradiation scenarios on healthy tissue response across diverse environmental conditions}
\label{Subsubsection: discussion NTCP}

Regarding the numerical application of the developed mechanistic NTCP model, in the previous section, we presented the main computational results obtained through the GSM$^2$-driven NTCP model. This \textit{in-silico} analysis highlights the key characteristics of the proposed mechanistic dose–response model and demonstrates its qualitative consistency with established experimental observations. In particular, we want to underline how this tool is capable of considering, from a physical point of view, various irradiation schemes, considering different radiation quality of the beam, partial irradiation experiments, and fractionation schemes; while, from a biological point of view, the tool can investigate different cell lines with different radiosensitivity, oxygen distribution, and macroscopic biological systems with characteristic volume effects. 

Since the mechanistic part originates from the GSM$^2$, the developed NTCP model can describe the biological effect of any radiation quality, i.e., particle type, energy, and LET, of the irradiation beam. This aspect emerges in Figures \ref{fig:NTCP_LET_Fr}~(a), \ref{fig:NTCP_p}, and \ref{fig:NTCP_C}, where we can observe the impact of radiation quality on the induction of a side effect for normal tissues. In detail, for a fixed organ type, characterized by cell line, volume effect, and oxygenation level, and for a given irradiation approach, the probability curve of the occurrence of a specific complication for that organ, specifically the tolerance dose $\mathrm{TD_{50}}$, shifts towards higher doses with decreasing LET values, as expected. This is a well-known effect, documented in several works in the literature such as \cite{SAAGER2018, HINTZ2022, KARGER2006}. This is due to increased biological effectiveness as a function of radiation quality, which is typically parameterized in terms of RBE. 
%As recently demonstrated in \cite{Bordieri2024}, the GSM$^2$ model can describe the different biological effects induced by different qualities of radiation at the cellular scale. In this work, we provide the first demonstration of the GSM$^2$ model's predictive capability in describing biological effects induced by various radiation qualities, even in complex biological systems on macroscopic scales. Moreover, because the model is grounded in microdosimetry, the macroscopic radiobiological endpoint is not derived from mean values assuming Poisson statistics. Rather, it results from explicitly accounting for stochastic, cell‑level fluctuations in energy deposition.

The second feature of the present model we wish to emphasize is its capability to describe different fractionation schemes. This is evident in Figure \ref{fig:NTCP_LET_Fr}~(b) in the case of a $100~\mathrm{MeV}$ proton beam, where, using the same set of biological parameters, the model predicts the well-known fractionation effect \cite{barendsen1982dose}, for three different numbers of fractions, namely 5, 15, and 30. In particular, the probability of radiation-induced complications in normal tissues significantly decreases as the number of fractions increases. This is because delivering the dose in a fractionated scheme allows cells more time to repair damage.

A key general point that emerges from most of the panels in Figures \ref{fig:NTCP_p}-\ref{fig:NTCP_C} is that the NTCP curves are not perfectly symmetric and sigmoidal as a function of dose, as typically assumed in most of the NTCP modeling works in literature \cite{kallman1992tumour, niemierko1993modeling, niemierko1999geud, lyman1985complication, kutcherburman1989ntcplkb, kutcher1991histogram}. This effect arises from a combination of factors, primarily linked to heterogeneity. Specifically, it results from the interplay between dose heterogeneity and cellular heterogeneity in terms of intrinsic radiosensitivity and oxygenation status. Each cell receives a distinct dose delivered by particles of different energies, and the associated biological response is governed by stochastic processes in damage induction and evolution. %These features are naturally captured by the GSM$^2$ model, which accounts for non‑Poissonian effects and enables a realistic description of variability at the microscopic scale. 
For this reason, the $\mathrm{TS}_{25-75}$ metric is used in the analyses reported in Figure \ref{fig:NTCP_index_p} rather than the standard $m_{50}$ metric, as the latter relies on an assumption of symmetry with respect to the inflection point, which dis not true for most of the cases considered.

A key mechanism through which tissue heterogeneity is incorporated into the GSM$^2$-driven TCP/NTCP framework is the assignment of different oxygenation levels to individual cells. This leads to cell-specific oxygen fixation of DNA damage~\cite{Scifoni2013}, which is modeled through the OER, as previously discussed. From Figures \ref{fig:NTCP_p}-\ref{fig:NTCP_C}, we can appreciate the impact of different oxygenation levels on the induction of a side effect. In particular, as expected, the probability of a normal tissue complication decreases due to reducing the environmental oxygen concentration for a fixed given dose. This effect is clearly illustrated in Figure \ref{fig:NTCP_index_p}, where we can see the shift of $\mathrm{TD_{50}}$ towards higher dose values for decreasing oxygen concentrations. In addition, Figure \ref{fig:NTCP_index_p} shows that the slope of the dose–response curve $m_{50}$, described by the derivative $\mathrm{TS}_{25-75}$, depends on the oxygen level, increasing with oxygenation and eventually reaching a plateau. Furthermore, the OER effect decreases with increasing radiation quality, according to equation \eqref{EQ:OER_LET_O2}. This effect emerges by comparing the dose-response curves for the two different radiation qualities, i.e., p $100~\mathrm{MeV}$ and $^{12}$C $280~\mathrm{MeV/u}$, reported in Figures \ref{fig:NTCP_p} and \ref{fig:NTCP_C}, respectively.

In addition, the mechanistic-driven NTCP model can exploit the impact of partial irradiation and volume effect on the manifestation of a side effect in healthy tissues. In detail, from Figures \ref{fig:NTCP_p}-\ref{fig:NTCP_C}, we can observe the combined action of partial irradiated volumes and tissue seriality. For example, fixing the irradiation approach, i.e., whole-organ or partial irradiation, the considered organ responds to the radiation in different ways, depending on its characteristic biological architecture. In particular, Figures \ref{fig:NTCP_p}-\ref{fig:NTCP_C} illustrate that the influence of tissue architecture is minimal for whole-organ irradiation. In contrast, Figures \ref{fig:NTCP_p}-\ref{fig:NTCP_C} highlight the emergence of volume effects in cases of partial organ irradiation, qualitatively aligning with widely recognized findings in the literature \cite{kutcher1991histogram, lyman1985complication}. This is because irradiating only a portion of the target volume, in the case of a parallel organ, reduces the probability of complications. Furthermore, the emergence of volume effects is emphasized by the $\mathrm{TD_{50}}$ of dose-response curves in Figure \ref{fig:NTCP_index_p}, which clearly depends on the percentage of irradiated volume for organs with low seriality ($s \rightarrow 0$), i.e. mostly parallel organs, and decreases as seriality increases, eventually becoming independent of volume for completely serial organs ($s=1$). Thus, a larger NTCP curve shift per unit reduction in partial volume (towards higher doses) reflects a greater volume effect, as emphasized in \cite{lyman1985complication, emami1991tolerance}. Lastly, from Figure \ref{fig:NTCP_index_p} we can appreciate the impact of organ seriality on the emergence of volume effects in terms of the bending of the NTCP curves, i.e., $m_{50}$. In detail, the derivative $\mathrm{TS}_{25-75}$, defined in equation \eqref{eq:TS50}, decreases significantly as the irradiated volume is reduced in the case of parallel organs, as typically observed experimentally \cite{kutcher1991histogram, burman1991fitting, emami1991tolerance}. Instead, this effect is attenuated for serial organs, as expected \cite{kutcher1991histogram, burman1991fitting, emami1991tolerance}. Therefore, the combined impact of prescribed dose and partial-volume irradiation is reflected in both the tolerance dose $\mathrm{TD_{50}}$ and the slope $\mathrm{TS}_{25-75}$ of the NTCP curve. For organs with intermediate seriality, changes in the slope are less pronounced than for parallel organs when the same partial volume is irradiated, because a relatively small increase in the maximum dose delivered to that portion of the organ is sufficient to produce the same level of toxicity. In contrast, the more parallel an organ is, the less sensitive it is to a local increase in dose per unit of partial volume irradiated. In these organs, the spared organ fraction can preserve its overall function, resulting in flatter NTCP curves with decreasing irradiated volume as a function of the prescribed dose. This behavior has been highlighted in other works using different NTCP models, such as in \cite{lyman1985complication, burman1991fitting}.

\subsection{Physical and environmental heterogeneities effects on tumor response}
\label{Subsubsection: discussion TCP}

To illustrate the potential of the proposed computational model in describing tumor response to particle therapy, we analyze the effect of different particles on 3D spheroids—aggregates of tumor cells that also exhibit oxygenation gradients. Three-dimensional spheroids are gaining increasing attention in experimental radiobiology, as they provide an \textit{in-vitro} system that more closely resembles \textit{in-vivo} tumor architecture compared to standard 2D clonogenic assays. A computational tool capable of reproducing the complex structure and microenvironment of 3D spheroids is therefore essential to fully exploit \textit{in-silico} analyses and to support the interpretation and design of experimental studies. This will enable us to calibrate and validate the model using clinically relevant endpoints beyond the conventional 2D clonogenic assay. Moreover, the same framework can be extended to calibration directly on clinical data.

One of the main advantage of the presented approach is that the model can take into account inter- and intra- cellular effects. In particular, our method allows us to consider, for example, specific spatial and temporal dose distributions, oxygenation gradients, ensembles of different cell lines, etc.

This part of the study focuses on how the oxygen distribution influences the response of the irradiated tissue. In particular, the results presented here were obtained by assuming a single cell type, but comparing an oxygenation gradient with a uniform distribution of oxygen in the spheroid volume, as reported in Figure \ref{fig:TCP_spheroid_O2}. In detail, Figure \ref{fig:TCP_spheroid_O2}~(d) highlights the importance of accounting for the actual oxygen distribution rather than its mean value. The impact of the oxygenation gradient is quantified by a shift of the TCP curve towards higher doses by a factor of 1.4 for the $100~\mathrm{MeV}$ proton beam irradiation, while 1.3 for the $80~\mathrm{MeV/u}$ carbon ion beam. This effect is due to the presence of hypoxic cells, which exhibit lower radiosensitivity compared with cells at physioxic oxygen levels. This effect is reflected not only in the $\mathrm{TD_{50}}$ but also in the slope $m_{50}$ of the TCP curves, and on the distribution of the simulated data points. In particular, for the oxygenation gradient condition, the TCP points exhibit a higher and significant stochastic variability, compared to the uniform oxygenation condition, where the TCP point follow a smooth trend, due to the different heterogeneity of the tissue oxygenation. Furthermore, this effect is further emphasized in carbon ion irradiation compared to proton irradiation, due to the inherent higher stochasticity in energy deposition compared to protons. As shown in Figure \ref{fig:TCP_spheroid_O2}~(c) through the cell survival results for the $100~\mathrm{MeV}$ proton beam, the reduced radiosensitivity in the oxygenation gradient setup becomes increasingly pronounced at higher dose levels. In particular, it becomes evident that the resulting cell-survival curve does not follow the typical linear--quadratic behavior; instead, it exhibits a pronounced deviation around 2\,Gy. This occurs because, at this dose, most of the sensitive cells have been eliminated, and the remaining population is predominantly composed of more resistant, that is, less oxygenated, cells. An interesting aspect is highlighted in Figure \ref{fig:TCP_spheroid_O2}~(b), which illustrates the impact of the two oxygen conditions on the initial distribution of $X$-lesions, for p $100~\mathrm{MeV}$. The sub-lethal DNA damage distribution in the case of uniform mean oxygenation is characterized by a symmetric shape around the average induced X-damage, according to equation \eqref{eq:kDNA}, after an absorbed dose of 6 Gy. Instead, for the oxygenation gradient case, the distribution of X-lesions loses the previously described symmetry due to the presence of cells with different oxygenation levels. In particular, the frequency of having fewer damages than the mean value increases because the effect of a reduced oxygenation level enhances the effect quantified by the OER, which is the so-called "oxygen fixation" effect \cite{Scifoni2013}. This different induction of sub-lethal damage at the cellular level underlies the different macroscopic response observed in terms of TCP between these two case studies.

The second part of this study focuses on investigating the role of cell line distributions in terms of biological response to particle therapy. Compared to the previously discussed setup, in this case we considered a spheroid made by two cell populations with different radiosensitivity, i.e., a sensitive and a resistant population, as illustrated in Figure \ref{fig:TCP_spheroid_co}~(a), and we assumed uniform oxygenation throughout the volume. In particular, Figure \ref{fig:TCP_spheroid_co}~(d) shows the impact of considering the actual cell type distribution, by comparing the co-culture with spheroids with a single radiosensitivity. As we can see from these TCP curves, the more the spheroid is composed of radiosensitive cells, the more the $\mathrm{TD_{50}}$ shifts toward lower doses, as expected. In particular, the shifts of the (58$\%$-42$\%$)/(90$\%$-10$\%$) co-culture TCP curves and the radiosensitive TCP curve from the radioresistant TCP curve are equal to a factor of 1.1/1.2 and 2.0, respectively. In addition, an interesting point is that in the case of co-culture, there is also a variation in the slope $m_{50}$ of the TCP curve, since at low doses the macroscopic response of the spheroid depends more on the surviving probability of radiosensitive cells, while at higher doses, the TCP curve depends more on the inactivation of radioresistant cells. This behavior is evident from the survival curves in Figures~\ref{fig:TCP_spheroid_co}(b), which deviate from the typical linear--quadratic trend, consistent with what was already observed in the analysis of oxygenation. At low doses, the survival probability of the mixed population closely follows that of the sensitive subpopulation, as highlighted in Figure ~\ref{fig:TCP_spheroid_co}(c). After a certain dose, around 1\,Gy in this case, the curve changes shape because most of the sensitive cells have been inactivated. Beyond this point, the overall survival curve follows the trend of the fully resistant population, though with a vertical shift due to the initial presence of the sensitive cells.

%\subsection{Radiation-induced response of the rat spinal cord}
%\label{Subsubsection: discussion comparison}

\subsection{Current limitations of GSM$^2$-driven NTCP and TCP model and future directions}
\label{Subsubsection: outlook}

The previous discussion highlights the generality and flexibility of the developed GSM$^2$-driven NTCP and TCP model. We discuss here the current limitations of the developed models, highlighting further future lines of research. 

The first point that we want to discuss is the computational cost of the model. This is not a problem for low doses or relatively small volumes, such as those considered in this study, with a cell density range of $10^4-10^5$ per $\mathrm{mm^3}$, as previously discussed. However, the situation could become more complex when considering very high doses, of several tens of Gy, or significantly larger target volumes, on the order of several $\mathrm{cm^3}$, arising from the single-cell resolution of the developed model. In this sense, although the code is highly parallelized and can therefore benefit from deployment on high‑performance computing clusters, the cell‑by‑cell description still imposes substantial memory demands when modeling extremely large populations. Moreover, in certain situations, such as irradiation with high‑energy protons, the general formulation produces results that are effectively equivalent to those obtained under Poisson statistics, making an overly general model unnecessary in those regimes \cite{cordoni2022cell}. To address this, we are developing a GPU-based approach for predicting NTCP and TCP curves under different experimental conditions, ensuring extremely fast computational times.

In this work, as discussed in the previous sections, we focused on the effects of protons, helium, and carbon ions irradiation on both healthy tissues and tumors, across a range of LET values relevant to current clinical practice. However, thanks to the flexibility of the proposed model, we could also extend this study to heavier ions with higher LET values, such as oxygen and neon ions, which are becoming increasingly relevant in the field of particle therapy. Such analyses will be investigated in future works.

%The main limitation of the present model is the biological description of organs with complex structures or intricate radiation-induced complications. In this study, we consider a single cell type, for simplicity, calibrating the organ response model on that cell line. This approximation may present limitations in the description of more sophisticated biological structures. In such cases, it becomes necessary to consider different cell types with specific biological characteristics, such as for the skin tissue. For this reason, we are focusing on reproducing the dose-response relationship for a relatively simple and well-established biological structure, such as the spinal cord. 
%However, it also demands a significantly more detailed understanding of both the irradiation process and the biological environment, often only partially characterized, along with additional parameters calibration.

A further important limitation of the current version of the model is the absence of repopulation dynamics, which are known to play a central role in both tumor control and normal‑tissue response, particularly over multi‑fraction or protracted treatments \cite{kim2005repopulation}. Because the present GSM$^2$-driven framework primarily focuses on radiation‑induced damage and its downstream biological consequences, it does not yet incorporate the mitotic regeneration processes that can significantly modify the dose–response relationship at the organ and tumor level. To address this gap, we are developing an intermediate agent‑based model explicitly describing cell proliferation, cell–cell interactions, and spatial constraints within the tissue microenvironment \cite{cordoni2026ab}. This additional layer will be coupled to the final TCP/NTCP model, enabling us to include repopulation in a mechanistically consistent manner. Such an extension will allow calibration against tumor‑growth experiments and longitudinal \textit{in‑vivo} measurements, thereby introducing a biologically critical mechanism and substantially improving the predictive realism of the model.

In addition, the model does not differentiate between the various DNA damage-mediated cell death pathways in this version. In this context, we are working to incorporate additional biological pathways, such as senescence and apoptosis, by integrating the proteins p21, p53, or TGF-$\beta$, which mediate these alternative cell-death processes \cite{schaue2012cytokines}. This will enable us to describe other biological endpoints, in particular radiation-mediated late effects, including, for instance, fibrosis being well known to be correlated to senescence. We are also considering the impact of ferroptosis, another form of cell death that does not involve DNA damage, by accounting for the role of membrane lipid damage. This extension will differentiate between DNA and lipids, enabling us to predict additional biological endpoints while also considering complex volume effects beyond the spinal cord, including radiation-induced complications for the skin or lungs. The idea is to explore possible interplays among radiation quality, fractionation scheme, and biological endpoint on the emergence of a side effect for PT. In this regard, we are already working on a more detailed description of the biological structure of the skin, considering its typical layered architecture, with each layer exhibiting specific chemical and biological characteristics \cite{rube2024radiation}. These final extensions are designed to capture recent experimental evidence on the biological response of complex systems exposed to ultra‑high dose‑rate (UHDR) irradiation \cite{favaudon2014ultrahigh, vozenin2019advantage}. In detail, developing the present predictive mechanistic model with single-cell resolution for the radiation response of macroscopic biological systems makes it an ideal candidate for the UHDR extension to investigate the FLASH effect. The main idea is to include the impact of radiation chemistry on the dose-response relationship of macroscopic \textit{in-vivo} biological systems, as already performed with developing the MS-GSM$^2$ model for the description of UHDR response at \textit{in-vitro} level in \cite{Battestini2023, battestini2024multiscale}. 

Lastly, we aim to extend the GSM$^2$-driven NTCP and TCP model to realistic clinical settings, quantifying treatment quality through mechanistic-grounded NTCP and TCP predictions. Integrating LET and dose distributions with environmental factors such as oxygenation, cell line, and organ type will inform the GSM$^2$-driven NTCP and TCP model to evaluate whole-organ response in clinical settings for PT. With this tool, we will assess the potential clinical impact of the discussed stochastic effects. Furthermore, the proposed approach will not only facilitate the evaluation of treatment plans in PT but also provide a foundation for creating innovative plan optimization methods that incorporate biological effects with a mechanistic basis, extending beyond the standard dosimetric constraints commonly used in clinical settings, such as in \cite{battestini2022including}, thanks to the potential integration of the GSM$^2$-driven NTCP and TCP model into a TPS. It is important to note that this final step represents an exceptionally challenging effort. Achieving a mechanistically grounded prediction of clinically relevant endpoints, such as NTCP and TCP, has been a longstanding goal in radiotherapy research, reflecting decades of effort to move beyond purely empirical or phenomenological descriptions. The integration of detailed radiobiological mechanisms with full clinical treatment‑planning information, therefore, constitutes not only a major extension of the GSM$^2$-driven framework but also a significant step toward a long‑sought objective in the field: a predictive, mechanistic model capable of informing and optimizing radiotherapy directly at the clinical level.

%% CONCLUSIONS
\section{Conclusions}
\label{4_Conclusions}
\counterwithin{figure}{section}
\setcounter{figure}{0}

In this work, we develop a novel mathematical dose-response model, expanding a radiation biophysical model, for mechanistic-driven predictions of possible healthy tissue complications and tumor control induced by particle beams across a broad range of physical and tissue environmental conditions. In detail, we integrated the GSM$^2$ into NTCP and TCP models, accounting for the specificities of energy deposition by particles for cellular response, to complex healthy organs. This approach will allow us to consider physical and environmental heterogeneities and stochastic effects that are typically neglected by standard NTCP and TCP models in the literature. 

Thanks to the generality and flexibility of the GSM$^2$-driven NTCP and TCP model, we assessed how different irradiation methods, in terms of radiation type, quality, and fraction schemes, and also tissue environmental conditions, affect the induction of side effects and explore the role of heterogeneities in oxygenation and cell type on tumor control. In particular, we showed the interplay between the radiation quality of the particle beam and the oxygenation level of the irradiated tissue in the manifestation of toxicities, which is modulated by the "oxygen fixation" effect. In addition, we underlined the impact of partial irradiation and tissue seriality on the emergence of volume effects in macroscopic biological systems for PT. Furthermore, we highlighted the importance of considering the actual oxygen and cell radiosensitivity distributions in spheroids. 

Lastly, the NTCP and TCP model will serve as a valuable tool for assessing treatment plan quality and predicting potential radiation-induced side effects, based on a unique mechanistic approach. In particular, the proposed model could be extended in the future to address more clinically relevant scenarios, with the aim of establishing a general framework for 3D dose distribution evaluation. This would enable radiobiological considerations in the plan evaluation, beyond standard purely dosimetric indices, for a more mechanistic-grounded investigation. Moreover, the developed NTCP model can be applied and adapted to a variety of organs at risk, biological endpoints, and cutting-edge irradiation techniques, including UHDR irradiation and mini-beams.

%% CONFLICTS OF INTEREST STATEMENT
\section*{Conflict of Interest Statement}
\label{6_Conflict_of_Interest_Statement}
\counterwithin{figure}{section}
\setcounter{figure}{0}
The authors have no relevant conflicts of interest to disclose.

%% FUNDINGS
\section*{Funding}
\label{7_funding}
\counterwithin{figure}{section}
\setcounter{figure}{0}
This work has been partially supported by the INFN CSN5 project FRIDA, and partially funded by the INFN CSN5 "Grant Giovani" project PROBE.

%% bib database file

\bibliographystyle{elsarticle-num} 
\bibliography{NTCP_GMS2}

@article{pfuhl2022comprehensive,
  title={Comprehensive comparison of local effect model IV predictions with the particle irradiation data ensemble},
  author={Pfuhl, Tabea and Friedrich, Thomas and Scholz, Michael},
  journal={Medical Physics},
  volume={49},
  number={1},
  pages={714--726},
  year={2022},
  publisher={Wiley Online Library}
}

@article{kundrat2020analytical,
  title={Analytical formulas representing track-structure simulations on DNA damage induced by protons and light ions at radiotherapy-relevant energies},
  author={Kundr{\'a}t, Pavel and Friedland, Werner and Becker, Janine and Eidem{\"u}ller, Markus and Ottolenghi, Andrea and Baiocco, Giorgio},
  journal={Scientific Reports},
  volume={10},
  number={1},
  pages={15775},
  year={2020},
  publisher={Nature Publishing Group UK London}
}

@article{cordoni2021generalized,
  title={Generalized stochastic microdosimetric model: The main formulation},
  author={Cordoni, F and Missiaggia, M and Attili, A and Welford, SM and Scifoni, E and La Tessa, C},
  journal={Physical Review E},
  volume={103},
  number={1},
  pages={012412},
  year={2021},
  publisher={APS},
    doi = "10.1103/PhysRevE.103.012412"
}

@article{battestini2022including,
  title={Including Volume Effects in Biological Treatment Plan Optimization for Carbon Ion Therapy: Generalized Equivalent Uniform Dose-Based Objective in TRiP98},
  author={Battestini, Marco and Schwarz, Marco and Kr{\"a}mer, Michael and Scifoni, Emanuele},
  journal={Frontiers in Oncology},
  volume={12},
  pages={555},
  year={2022},
  publisher={Frontiers},
 doi={https://doi.org/10.3389/fonc.2022.826414}
}

@article{missiaggia2020microdosimetric,
  title={Microdosimetric measurements as a tool to assess potential in-field and out-of-field toxicity regions in proton therapy},
  author={Missiaggia, M and Cartechini, G and Scifoni, E and Rovituso, M and Tommasino, F and Verroi, E and Durante, M and La Tessa, C},
  journal={Physics in Medicine \& Biology},
  volume={65},
  number={24},
  pages={245024},
  year={2020},
  publisher={IOP Publishing}
}

@article{missiaggia2021novel,
  title={A novel hybrid microdosimeter for radiation field characterization based on the tissue equivalent proportional counter detector and low gain avalanche detectors tracker: a feasibility study},
  author={Missiaggia, M and Pierobon, E and Castelluzzo, M and Perinelli, A and Cordoni, F and Centis Vignali, M and Borghi, G and Bellinzona, EV and Scifoni, E and Tommasino, F and others},
  journal={Frontiers in Physics},
  volume={8},
  pages={578444},
  year={2021},
  publisher={Frontiers Media SA}
}

@article{missiaggia2023investigation,
  title={Investigation of in-field and out-of-field radiation quality with microdosimetry and its impact on relative biological effectiveness in proton therapy},
  author={Missiaggia, Marta and Cartechini, Giorgio and Tommasino, Francesco and Scifoni, Emanuele and La Tessa, Chiara},
  journal={International Journal of Radiation Oncology* Biology* Physics},
  volume={115},
  number={5},
  pages={1269--1282},
  year={2023},
  publisher={Elsevier}
}

@article{cordoni2023emergence,
  title={On the Emergence of the Deviation from a Poisson Law in Stochastic Mathematical Models for Radiation-Induced DNA Damage: A System Size Expansion},
  author={Cordoni, Francesco Giuseppe},
  journal={Entropy},
  volume={25},
  number={9},
  pages={1322},
  year={2023},
  publisher={MDPI}
}

@article{cordoni2022multiple,
  title={Multiple levels of stochasticity accounted for in different radiation biophysical models: from physics to biology},
  author={Cordoni, Francesco G and Missiaggia, Marta and La Tessa, Chiara and Scifoni, Emanuele},
  journal={International Journal of Radiation Biology},
  pages={1--16},
  year={2022},
  publisher={Taylor \& Francis},
    doi = "https://doi.org/10.1080/09553002.2023.2146230"
}

@article{cordoni2022cell,
  title={Cell Survival Computation via the Generalized Stochastic Microdosimetric Model (GSM2); Part I: The Theoretical Framework},
  author={Cordoni, Francesco G and Missiaggia, Marta and Scifoni, Emanuele and La Tessa, Chiara},
  journal={Radiation Research},
  volume={197},
  number={3},
  pages={218--232},
  year={2022},
  publisher={Radiation Research Society},
    doi ="https://doi.org/10.1667/RADE-21-00098.1"
}

@article{favaudon2014ultrahigh,
  title={Ultrahigh dose-rate FLASH irradiation increases the differential response between normal and tumor tissue in mice},
  author={Favaudon, Vincent and Caplier, Laura and Monceau, Virginie and Pouzoulet, Fr{\'e}d{\'e}ric and Sayarath, Mano and Fouillade, Charles and Poupon, Marie-France and Brito, Isabel and Hup{\'e}, Philippe and Bourhis, Jean and others},
  journal={Science translational medicine},
  volume={6},
  number={245},
  pages={245ra93--245ra93},
  year={2014},
  publisher={American Association for the Advancement of Science}
}

@article{vozenin2019advantage,
  title={The Advantage of FLASH Radiotherapy Confirmed in Mini-pig and Cat-cancer PatientsThe Advantage of Flash Radiotherapy},
  author={Vozenin, Marie-Catherine and De Fornel, Pauline and Petersson, Kristoffer and Favaudon, Vincent and Jaccard, Maud and Germond, Jean-Fran{\c{c}}ois and Petit, Benoit and Burki, Marco and Ferrand, Gis{\`e}le and Patin, David and others},
  journal={Clinical Cancer Research},
  volume={25},
  number={1},
  pages={35--42},
  year={2019},
  publisher={AACR}
}

@article{kase2007biophysical,
  title={Biophysical calculation of cell survival probabilities using amorphous track structure models for heavy-ion irradiation},
  author={Kase, Yuki and Kanai, Tatsuaki and Matsufuji, Naruhiro and Furusawa, Yoshiya and Els{\"a}sser, Thilo and Scholz, Michael},
  journal={Physics in Medicine \& Biology},
  volume={53},
  number={1},
  pages={37},
  year={2007},
  publisher={IOP Publishing},
    doi = "10.1088/0031-9155/53/1/003"
}

@book{Ros,
  title={Microdosimetry and its Applications},
  author={Zaider, M and Rossi, By Harald H and Zaider, Marco},
  year={1996},
  publisher={Springer}
}

@Article{Bellinzona2021,
	author = "Bellinzona, V. E. and Cordoni, F. and Missiaggia, M. and Tommasino, F. and Scifoni, E. and La Tessa, C. and Attili, A.",
	title = "Linking Microdosimetric Measurements to Biological Effectiveness in Ion Beam Therapy: A Review of Theoretical Aspects of MKM and Other Models",
    journal = "Frontiers in Physics",
	volume = "8",
	number = "",
	pages = "",
	year = "2021",
	doi = "https://doi.org/10.3389/fphy.2020.578492",
}

@article{Battestini2023,
  
AUTHOR={Battestini, Marco and Missiaggia, Marta and Attili, Andrea and Tommasino, Francesco and La Tessa, Chiara and Cordoni, Francesco G. and Scifoni, Emanuele},   
	 
TITLE={Across the stages: a multiscale extension of the generalized stochastic microdosimetric model (MS-GSM2) to include the ultra-high dose rate},      
	
JOURNAL={Frontiers in Physics},      
	
VOLUME={11},           
	
YEAR={2023},      
	  
URL={https://www.frontiersin.org/articles/10.3389/fphy.2023.1274064},       
	
DOI={10.3389/fphy.2023.1274064},      
	
ISSN={2296-424X},   
   
}

@article{missiaggia2022cell,
    author = {Missiaggia, M. and Cordoni, F. G. and Scifoni, E. and Tessa, C. La},
    title = "{Cell Survival Computation via the Generalized Stochastic Microdosimetric Model (GSM2); Part II: Numerical Results}",
    journal = {Radiation Research},
    volume = {201},
    number = {2},
    pages = {104-114},
    year = {2024},
    month = {01},
    issn = {0033-7587},
    doi = {10.1667/RADE-22-00025.1.S1},
    url = {https://doi.org/10.1667/RADE-22-00025.1.S1},
}

@article{Scifoni2013,
doi = {10.1088/0031-9155/58/11/3871},
url = {https://dx.doi.org/10.1088/0031-9155/58/11/3871},
year = {2013},
month = {may},
publisher = {IOP Publishing},
volume = {58},
number = {11},
pages = {3871},
author = {E Scifoni and W Tinganelli and W K Weyrather and M Durante and A Maier and M Krämer},
title = {Including oxygen enhancement ratio in ion beam treatment planning: model implementation and experimental verification},
journal = {Physics in Medicine \& Biology},
}

@article{Bordieri2024,
author={Bordieri, Giulio and Missiaggia, Marta and Cartechini, Giorgio and Battestini, Marco and Bronk, Lawrence and Guan, Fada and Grosshans, David R and Rai, Priyamvada and Scifoni, Emanuele and La Tessa, Chiara and Lattanzi, Gianluca and Cordoni, Francesco G},
title={Validation of the generalized stochastic microdosimetric model (GSM2) over a broad range of LET and particle beam type: a unique model for accurate description of (therapy relevant) radiation qualities},
journal={Physics in Medicine \& Biology},
url={http://iopscience.iop.org/article/10.1088/1361-6560/ad9dab},
year={2024},
}

@article{SAAGER2018,
title = {Determination of the proton RBE in the rat spinal cord: Is there an increase towards the end of the spread-out Bragg peak?},
journal = {Radiotherapy and Oncology},
volume = {128},
number = {1},
pages = {115-120},
year = {2018},
issn = {0167-8140},
doi = {https://doi.org/10.1016/j.radonc.2018.03.002},
url = {https://www.sciencedirect.com/science/article/pii/S0167814018301348},
author = {Maria Saager and Peter Peschke and Stephan Brons and Jürgen Debus and Christian P. Karger},
}

@article{HINTZ2022,
title = {Relative biological effectiveness of single and split helium ion doses in the rat spinal cord increases strongly with linear energy transfer},
journal = {Radiotherapy and Oncology},
volume = {170},
pages = {224-230},
year = {2022},
issn = {0167-8140},
doi = {https://doi.org/10.1016/j.radonc.2022.03.017},
url = {https://www.sciencedirect.com/science/article/pii/S0167814022001554},
author = {Lisa Hintz and Christin Glowa and Maria Saager and Rosemarie Euler-Lange and Peter Peschke and Stephan Brons and Rebecca Grün and Michael Scholz and Stewart Mein and Andrea Mairani and Jürgen Debus and Christian P. Karger},
}

@article{KARGER2006,
title = {Radiation tolerance of the rat spinal cord after 6 and 18 fractions of photons and carbon ions: Experimental results and clinical implications},
journal = {International Journal of Radiation Oncology*Biology*Physics},
volume = {66},
number = {5},
pages = {1488-1497},
year = {2006},
issn = {0360-3016},
doi = {https://doi.org/10.1016/j.ijrobp.2006.08.045},
url = {https://www.sciencedirect.com/science/article/pii/S0360301606028082},
author = {Christian P. Karger and Peter Peschke and Rita Sanchez-Brandelik and Michael Scholz and Jürgen Debus},
}

@article{epp1972radiosensitivity,
  title={The radiosensitivity of cultured mammalian cells exposed to single high intensity pulses of electrons in various concentrations of oxygen},
  author={Epp, Edward R and Weiss, Herbert and Djordjevic, Bozidar and Santomasso, Ann},
  journal={Radiation research},
  volume={52},
  number={2},
  pages={324--332},
  year={1972},
  publisher={Academic Press, Inc.},
doi={https://doi.org/10.2307/3573572}
}

@article{battestini2024multiscale,
  title={A multiscale radiation biophysical stochastic model describing the cell survival response at ultra-high dose rate under different oxygenations and radiation qualities},
  author={Battestini, Marco and Missiaggia, Marta and Bolzoni, Sara and Cordoni, Francesco G and Scifoni, Emanuele},
  journal={Radiotherapy and Oncology},
  year={2025},
  publisher={Elsevier}
}

@article{scholz2003effects,
  title={Effects of ion radiation on cells and tissues},
  author={Scholz, Michael},
  journal={Radiation effects on polymers for biological use},
  pages={95--155},
  year={2003},
  publisher={Springer},
doi={https://doi.org/10.1007/3-540-45668-6_4}
}

@article{Missiaggia2024,
    author = {Missiaggia, M. and Cordoni, F. G. and Scifoni, E. and Tessa, C. La},
    title = "{Cell Survival Computation via the Generalized Stochastic Microdosimetric Model (GSM2); Part II: Numerical Results}",
    journal = {Radiation Research},
    volume = {201},
    number = {2},
    pages = {104-114},
    year = {2024},
    month = {01},
    issn = {0033-7587},
    doi = {10.1667/RADE-22-00025.1.S1},
    url = {https://doi.org/10.1667/RADE-22-00025.1.S1},
}

@article{bidanta2023functional,
  title={Functional Tissue Units in the Human Reference Atlas},
  author={Bidanta, Supriya and B{\"o}rner, Katy and Herr, Bruce W and Nagy, Marcell and Gustilo, Katherine S and Bajema, Rachel and Maier, Libby and Molontay, Roland and Weber, Griffin and others},
  journal={bioRxiv},
  year={2023},
  publisher={Cold Spring Harbor Laboratory Preprints}
}

@article{niemierko1993modeling,
  title={Modeling of normal tissue response to radiation: the critical volume model},
  author={Niemierko, Andrzej and Goitein, Michael},
  journal={International Journal of Radiation Oncology* Biology* Physics},
  volume={25},
  number={1},
  pages={135--145},
  year={1993},
  publisher={Elsevier}
}

@article{kallman1992tumour,
  title={Tumour and normal tissue responses to fractionated non-uniform dose delivery},
  author={K{\"a}llman, Per and {\AA}gren, Anders and Brahme, Anders},
  journal={International journal of radiation biology},
  volume={62},
  number={2},
  pages={249--262},
  year={1992},
  publisher={Taylor \& Francis}
}

@article{Cometto2014,
author = {Cometto, A. and Russo, G. and Bourhaleb, F. and Milian, F. M. and Giordanengo, S. and Marchetto, F. and Cirio, R. and Attili, A.},
doi = {10.1088/0031-9155/59/23/7393},
issn = {13616560},
journal = {Phys. Med. Biol.},
number = {23},
pages = {7393--7417},
pmid = {25386876},
title = {{Direct evaluation of radiobiological parameters from clinical data in the case of ion beam therapy: An alternative approach to the relative biological effectiveness}},
volume = {59},
year = {2014}
}

@article{barendsen1982dose,
  title={Dose fractionation, dose rate and iso-effect relationships for normal tissue responses},
  author={Barendsen, GW},
  journal={International Journal of Radiation Oncology* Biology* Physics},
  volume={8},
  number={11},
  pages={1981--1997},
  year={1982},
  publisher={Elsevier},
doi={https://doi.org/10.1016/0360-3016(82)90459-X}
}

@article{kutcher1991histogram,
  title={Histogram reduction method for calculating complication probabilities for three-dimensional treatment planning evaluations},
  author={Kutcher, GJ and Burman, C and Brewster, L and Goitein, M and Mohan, R},
  journal={International Journal of Radiation Oncology* Biology* Physics},
  volume={21},
  number={1},
  pages={137--146},
  year={1991},
  publisher={Elsevier},
doi={https://doi.org/10.1016/0360-3016(91)90173-2}
}

@article{lyman1985complication,
  title={Complication probability as assessed from dose-volume histograms},
  author={Lyman, John T},
  journal={Radiation research},
  volume={104},
  number={2s},
  pages={S13--S19},
  year={1985},
  publisher={Academic Press, Inc.},
doi={https://doi.org/10.2307/3576626}
}

@article{rube2024radiation,
  title={Radiation dermatitis: radiation-induced effects on the structural and immunological barrier function of the epidermis},
  author={R{\"u}be, Claudia E and Freyter, Benjamin M and Tewary, Gargi and Roemer, Klaus and Hecht, Markus and R{\"u}be, Christian},
  journal={International Journal of Molecular Sciences},
  volume={25},
  number={6},
  pages={3320},
  year={2024},
  publisher={MDPI},
doi={https://doi.org/10.3390/ijms25063320}
}

@article{withers1988fsu,
  title={Treatment volume and tissue tolerance},
  author={Rodney Withers, H and Taylor, J M G and Maciejewski, B},
  journal={International Journal of Radiation Oncology - Biology - Physics},
  volume={14},
  pages={751-759},
  year={1988},
  publisher={Elsevier},
doi={https://doi.org/10.1016/0360-3016(88)90098-3}
}

@incollection{scholz2006dose,
  title={Dose response of biological systems to low- and high-LET radiation},
  author={Scholz, M},
  booktitle={Microdosimetric response of physical and biological systems to low- and high-LET radiations},
  pages={1-73},
  year={2006},
  publisher={Elsevier},
doi={https://doi.org/10.1016/B978-044451643-5/50013-7}
}

@article{kutcherburman1989ntcplkb,
  title={Calculation of complication probability factors for non-uniform normal tissue irradiation: the effective volume method},
  author={Kutcher, G J and Burman, C},
  journal={International Journal of Radiation Oncology - Biology - Physics},
  volume={16},
  pages={1623-1630},
  year={1989},
doi={https://doi.org/10.1016/0360-3016(89)90972-3}
}

@article{fellin2009clinical,
  title={Clinical and dosimetric predictors of late rectal toxicity after conformal radiation for localized prostate cancer: results of a large multicenter observational study},
  author={Fellin, Gianni and Fiorino, Claudio and Rancati, Tiziana and Vavassori, Vittorio and Baccolini, Micaela and Bianchi, Carla and Cagna, Emanuela and Gabriele, Pietro and Mauro, Floranna and Menegotti, Loris and others},
  journal={Radiotherapy and Oncology},
  volume={93},
  number={2},
  pages={197--202},
  year={2009},
  publisher={Elsevier},
doi={https://doi.org/10.1016/j.radonc.2009.09.004}
}

@article{niemierko1999geud,
  title={A generalized concept of Equivalent Uniform Dose (EUD)},
  author={Niemierko, A},
  journal={Medical Physics},
  volume={26},
  number={6},
  pages={1100},
  year={1999},
  }

@article{kiefer1986model,
  title={A model of ion track structure based on classical collision dynamics (radiobiology application)},
  author={Kiefer, Jiirgen and Straaten, Hermann},
  journal={Physics in Medicine \& Biology},
  volume={31},
  number={11},
  pages={1201},
  year={1986},
  publisher={IOP Publishing},
doi={10.1088/0031-9155/31/11/002}
}

@article{bezanson2017julia,
  title={Julia: A fresh approach to numerical computing},
  author={Bezanson, Jeff and Edelman, Alan and Karpinski, Stefan and Shah, Viral B},
  journal={SIAM review},
  volume={59},
  number={1},
  pages={65--98},
  year={2017},
  publisher={SIAM},
doi={https://doi.org/10.1137/141000671}
}

@article{bordieri2025integrating,
  title={Integrating nano-and micrometer-scale energy deposition models for mechanistic prediction of radiation-induced DNA damage and cell survival},
  author={Bordieri, Giulio and Missiaggia, Marta and Lattanzi, Gianluca and Villagrasa, Carmen and Perrot, Yann and Cordoni, Francesco G},
  journal={Computers in Biology and Medicine},
  volume={199},
  pages={111330},
  year={2025},
  publisher={Elsevier}
}

@article{chatterjee1976microdosimetric,
  title={Microdosimetric structure of heavy ion tracks in tissue},
  author={Chatterjee, A and Schaefer, HJ},
  journal={Radiation and environmental biophysics},
  volume={13},
  number={3},
  pages={215--227},
  year={1976},
  publisher={Springer}
}

@techreport{ICRU2016,
  author = {ICRU},
  title = {ICRU Report 90: Key Data for Ionizing-Radiation Dosimetry: Measurement Standards and Applications},
  institution = {International Commission on Radiation Units and Measurements},
  year = {2016}
}

@article{schaue2012cytokines,
  title={Cytokines in radiobiological responses: a review},
  author={Schaue, D{\"o}rthe and Kachikwu, Evelyn L and McBride, William H},
  journal={Radiation research},
  volume={178},
  number={6},
  pages={505--523},
  year={2012},
  publisher={The Radiation Research Society},
doi={https://doi.org/10.1667/RR3031.1}
}

@article{webb1993model,
  title={A model for calculating tumour control probability in radiotherapy including the effects of inhomogeneous distributions of dose and clonogenic cell density},
  author={Webb, Steve and Nahum, Alan Effraim},
  journal={Physics in medicine \& biology},
  volume={38},
  number={6},
  pages={653},
  year={1993},
  publisher={IOP Publishing},
  doi={10.1088/0031-9155/38/6/001}
}

@article{niemierko1993implementation,
  title={Implementation of a model for estimating tumor control probability for an inhomogeneously irradiated tumor},
  author={Niemierko, Andrzej and Goitein, Michael},
  journal={Radiotherapy and Oncology},
  volume={29},
  number={2},
  pages={140--147},
  year={1993},
  publisher={Elsevier},
  doi={https://doi.org/10.1016/0167-8140(93)90239-5}
}

@article{bidanta2025functional,
  title={Functional tissue units in the Human Reference Atlas},
  author={Bidanta, Supriya and B{\"o}rner, Katy and Herr Ii, Bruce W and Quardokus, Ellen M and Nagy, Marcell and Gustilo, Katherine S and Bajema, Rachel and Maier, Elizabeth and Molontay, Roland and Weber, Griffin M},
  journal={Nature Communications},
  volume={16},
  number={1},
  pages={1526},
  year={2025},
  publisher={Nature Publishing Group UK London},
  doi={https://doi.org/10.1038/s41467-024-54591-6}
}

@article{withers1988treatment,
  title={Treatment volume and tissue tolerance},
  author={Withers, H Rodney and Taylor, Jeremy MG and Maciejewski, Boguslaw},
  journal={International Journal of Radiation Oncology* Biology* Physics},
  volume={14},
  number={4},
  pages={751--759},
  year={1988},
  publisher={Elsevier},
  doi={https://doi.org/10.1016/0360-3016(88)90098-3}
}

@article{emami1991tolerance,
  title={Tolerance of normal tissue to therapeutic irradiation},
  author={Emami, Bahman and Lyman, J and Brown, A and Cola, L and Goitein, M and Munzenrider, JE and Shank, B and Solin, LJ and Wesson, M},
  journal={International Journal of Radiation Oncology* Biology* Physics},
  volume={21},
  number={1},
  pages={109--122},
  year={1991},
  publisher={Elsevier}, 
  doi={https://doi.org/10.1016/0360-3016(91)90171-Y}
}

@article{burman1991fitting,
  title={Fitting of normal tissue tolerance data to an analytic function},
  author={Burman, Chandra and Kutcher, GJ and Emami, B and Goitein, M},
  journal={International Journal of Radiation Oncology* Biology* Physics},
  volume={21},
  number={1},
  pages={123--135},
  year={1991},
  publisher={Elsevier},
  doi={https://doi.org/10.1016/0360-3016(91)90172-Z}
}

@article{sinclair1968cyclic,
  title={Cyclic x-ray responses in mammalian cells in vitro},
  author={Sinclair, Warren K},
  journal={Radiation research},
  volume={33},
  number={3},
  pages={620--643},
  year={1968},
  publisher={Academic Press, Inc.}
}

@article{kim2005repopulation,
  title={Repopulation of cancer cells during therapy: an important cause of treatment failure},
  author={Kim, John J and Tannock, Ian F},
  journal={Nature Reviews Cancer},
  volume={5},
  number={7},
  pages={516--525},
  year={2005},
  publisher={Nature Publishing Group UK London}
}

@article{citrin2026effects,
  title={Effects of Radiotherapy in Normal Tissue},
  author={Citrin, Deborah E and Timmerman, Robert D},
  journal={New England Journal of Medicine},
  volume={394},
  number={10},
  pages={996--1009},
  year={2026},
  publisher={Mass Medical Soc},
  doi={10.1056/NEJMra2506017}
}

@article{de2019radiotherapy,
  title={Radiotherapy toxicity},
  author={De Ruysscher, Dirk and Niedermann, Gabriele and Burnet, Neil G and Siva, Shankar and Lee, Anne WM and Hegi-Johnson, Fiona},
  journal={Nature reviews Disease primers},
  volume={5},
  number={1},
  pages={13},
  year={2019},
  publisher={Nature Publishing Group UK London},
  doi={https://doi.org/10.1038/s41572-019-0064-5}
}

@article{cordoni2026ab,
  title={A stochastic agent-based extension of the GSM2 model for particle therapy: cell-cycle dynamics, dose-rate dependence, and fractionation effects},
  author={Cordoni, Francesco G. and Battestini, Marco and Missiaggia, Marta},
  note      = {Submitted},
  year      = {2026}
}

\end{document}